\documentclass[12pt]{article}
\usepackage{latexsym}
\usepackage{epsfig,amssymb,euscript,slashed}
\usepackage[linktocpage]{hyperref}
\usepackage{amsmath}
\usepackage[nosort]{cite}
\usepackage{array,calc,epsfig}
\usepackage{bbm}
\usepackage{fancybox}

\usepackage[left=1.5cm,right=1.5cm,top=3.5cm,bottom=3.5cm]{geometry}

\DeclareMathOperator\arctanh{arctanh}

\numberwithin{equation}{section} 
\usepackage[]{graphicx}
\usepackage{color}

\begin{document}
\font\cmss=cmss10 \font\cmsss=cmss10 at 7pt


\hfill
	\vspace{18pt}
	\begin{center}
		{\Large 
			\textbf{The geometry of large charge multi-traces in $\mathcal{N}=4$ SYM}}
		
	\end{center}

	\vspace{8pt}
	\begin{center}
		{\textsl{Stefano Giusto and Alessandro Rosso}}
		
		\vspace{1cm}
				
		\textit{\small Dipartimento di Fisica,  Universit\`a di Genova, Via Dodecaneso 33, 16146, Genoa, Italy.} \\  \vspace{6pt}
		
		\textit{\small I.N.F.N. Sezione di Genova,
			Via Dodecaneso 33, 16146, Genoa, Italy.}\\
				
	\end{center}

	\vspace{12pt}
	
	\begin{center}
		\textbf{Abstract}
	\end{center}
	
	\vspace{4pt} {\small
		\noindent 
		We construct a one-parameter family of half-BPS solutions of type IIB supergravity using a consistent truncation to gauged five-dimensional supergravity. For small values of the parameter, the solution reduces to the linear perturbation of AdS$_5\times S^5$ dual to the chiral primary operator in the stress-tensor multiplet, and we give evidence that the geometry is regular and asymptotes AdS in a normalisable way for arbitrarily large values of the parameter. We conjecture that the solution is the gravity dual of a ``heavy" multi-trace operator in $\mathcal{N}=4$ $SU(N)$ Super Yang-Mills made by $p$ copies of the stress-tensor chiral primary operator, with $p$ of order $N^2$ in the large $N$ limit. We perform some holographic checks supporting this duality map.}

		\vspace{1cm}

		\thispagestyle{empty}

		\vfill
		\vskip 5.mm
		\hrule width 5.cm
		\vskip 2.mm
		{
			\noindent {\scriptsize e-mails: stefano.giusto@ge.infn.it, alessandro.rosso@ge.infn.it}
		}

		\setcounter{footnote}{0}
		\setcounter{page}{0}

		
		\baselineskip=17pt
		\parskip=5pt
		
		\newpage

\section{Introduction}
The states of a holographic CFT admit a gravitational description whose qualitative nature depends on how large the conformal dimension, $\Delta$, is. In the classical limit of large central charge,  $c$, the simplest and best understood regime is that of a finite conformal dimension that does not scale with $c$. Such ``light" states are dual to supergravity fluctuations around the AdS vacuum if they are supersymmetric, or, more generally, to string excitations. At the opposite end\footnote{Intermediate regimes include states with $\Delta\sim c^{1/4}$, corresponding to strings in the pp-wave background, and wrapped D-branes or giant gravitons, which provide the dual description of states with $\Delta\sim c^{1/2}$.} are ``heavy" states whose conformal dimension grows like $c$ when $c\to \infty$, which, in the limit of large 't Hooft coupling, are described by non-linear solutions of supergravity that are regular, horizonless and asymptote AdS at large distances. 

One of the reasons why this class of heavy states is interesting is because they can be identified with the microstates of black holes, at least in those cases in which their degeneracy grows sufficiently fast with their charges, which typically happens if they do not preserve too many supercharges. In more supersymmetric cases, the ensemble of heavy states -- but not the individual pure states -- is usually described by a singular supergravity solution which can be interpreted as a degenerate black hole with vanishing horizon area (examples of such singular solutions are the two-charge susy black hole \cite{Sen:1995in} and the half-BPS superstar \cite{Myers:2001aq}). It is the interest in black holes that has motivated most of the efforts in constructing the map between heavy states and regular supergravity solutions, generally quite a difficult task. 

In this endeavour, the most developed subfield is the supergravity construction of the microstates of the Strominger-Vafa black hole \cite{Strominger:1996sh}. If one restricts to the near-horizon region dual to a decoupled CFT, these microstates are represented by asymptotically AdS$_3\times S^3\times \mathcal{M}$ geometries, with $\mathcal{M}$ a 4D compact Ricci flat space, which usually plays only a spectator role. Not surprisingly, the first supergravity solutions to be found have been the ones with the most supersymmetries: the Lunin-Mathur geometries \cite{Lunin:2001jy,Kanitscheider:2007wq} dual to half-BPS heavy states preserving 16 superconformal charges in the decoupling limit and representing the microstates of the ``small" two-charge black hole. Studying the action on these geometries of the symmetries generating the CFT chiral algebra has lead to the construction of some of the quarter-BPS microstates of the Strominger-Vafa black hole \cite{Bena:2015bea,Bena:2016agb,Bena:2016ypk,Bena:2017xbt,Ceplak:2018pws,Heidmann:2019xrd} (also known as superstrata \cite{Shigemori:2020yuo}) and also of some non-supersymmetric states \cite{Ganchev:2021pgs,Ganchev:2021ewa,Ganchev:2023nef}. Despite being associated with the degenerate limit of a black hole, already the half-BPS solutions have contributed significant insights to holography. By using as probes the protected expectation values of light chiral primary operators (CPO's) in the heavy half-BPS states, one has formulated an impressively detailed map between CFT and gravity \cite{Kanitscheider:2006zf,Giusto:2015dfa,Giusto:2019qig,Rawash:2021pik}; moreover, from the study of the linear fluctuations around the microstate geometries, one has derived the strong coupling limit of the heavy-heavy-light-light (HHLL) correlator \cite{Galliani:2017jlg,Bombini:2017sge}, together with its light-light-light-light (LLLL) limits, including both single and multi-trace operators \cite{Giusto:2018ovt,Giusto:2019pxc,Giusto:2020neo,Ceplak:2021wzz}. 

Some of the progress in this AdS$_3$ holographic context has been hampered by the complexity of the CFT dual to the Strominger-Vafa black hole, the so called D1-D5 CFT. Already in the free limit, in which the CFT reduces to an orbifold sigma-model, field theory calculations hinge on some non-trivial technology \cite{Lunin:2000yv,Lunin:2001pw}, but the real stumbling block is the extension to the interacting theory, which is necessary to make contact with the gravity regime. At a generic point in its moduli space, the D1-D5 CFT does not have a known lagrangian description and computations can be done only perturbatively close to the free point by means of conformal perturbation theory (see for example \cite{Carson:2014ena}).

In this respect, a more convenient holographic model is represented by $\mathcal{N}=4$ $SU(N)$ Super Yang-Mills (SYM), whose states are dual to asymptotically AdS$_5\times S^5$ geometries. Despite this being probably the best studied interacting field theory, our supergravity understanding of its heavy states is surprisingly quite undeveloped and essentially\footnote{Some results on 1/4 and 1/8 BPS microstate geometries have been obtained; see for example \cite{jia2023interior}, for a very recent development, and the references therein for earlier attempts.} limited to the most supersymmetric, i.e. half-BPS, states, whose dual geometries have been known for quite some time \cite{Lin:2004nb}. These solutions, known as LLM geometries, are completely encoded in a droplet configuration drawn on a plane, which determines, via an integral formula, all the geometric data. Though the general integral expression allows to establish some important properties of the geometries, like their regularity or their asymptotic behaviour and global conserved charges, for a more detailed holographic analysis it is often useful to know the solution explicitly, i.e. to compute the integrals associated to some droplet configuration. Beside the simplest case of a circular droplet, which corresponds to the pure AdS$_5\times S^5$ vacuum, this is a difficult, or outright unfeasible, task for more generic droplets. For this reason most studies of the LLM geometries has focused on configurations with an extra $U(1)$ symmetry, like a central disk droplet with concentric rings. 

In the present article we would like to revisit the supergravity description of half-BPS states of $\mathcal{N}=4$ SYM for several reasons. Firstly, the map between LLM geometries and half-BPS operators is natural in the fermionic description of the latter \cite{Corley:2001zk,Berenstein:2004kk}: the fermions correspond to the eigenvalues of the $N$ by $N$ matrix that makes up the operators and the droplet that encodes the geometry describes the phase-space distribution of the fermions. Some holographic computations are however more naturally expressed in the multi-trace representations of the supersymmetric operators, where one expresses the heavy operators as the product of a large number (of order $N^2$) of  light multi-tsingle-trace CPO's. We focus here on the simplest such multi-traces formed by $p\sim N^2$ copies of identical single-trace constituents, and in particular we consider as elementary constituent the lightest CPO, that of dimension $\Delta=2$. We might refer to this class of operators as monochromatic multi-traces. It would be helpful to know the supergravity duals of these operators: this question, for example, was recently asked in \cite{Paul:2023rka}, since the knowledge of these geometries would allow, among other things, the holographic computation of the associated HHLL correlators and the comparison with the results obtained by other methods -- we have in mind, for example, the recent progress in the computation of integrated $\mathcal{N}=4$ correlators with heavy states \cite{Chester:2020dja,Brown:2023why}. 

Within the LLM framework, the problem is to identify the droplet configuration dual to the monochromatic operators. Since the R-charge carried by these operators breaks the extra $U(1)$ symmetry, it is clear that the relevant geometry does not fall into the class of the much studied concentric rings. In the regime in which $p/N^2$ is small, and thus the multi-trace approximates a light CPO, the answer to this question was already contained in the original work of LLM \cite{Lin:2004nb}, where it was suggested that the relevant phase-space configuration is a fluctuation of the circular droplet whose wavelength is related with the dimension of the light operator. This expectation was verified in \cite{Grant:2005qc} by showing that at linear order in $p/N^2$ the LLM geometry following from the above deformed droplet is diffeomorphic to the linear deformation of AdS$_5\times S^5$ dual to the light CPO \cite{Kim:1985ez}. Our concern here is the non-linear extension of this solution. A natural conjecture would be that monochromatic multi-traces are dual to monochromatic ripple deformations of the disk, i.e. to droplets whose boundary is described in polar coordinates by a curve of the form
\begin{equation}\label{eq:monodrop}
r(\phi) = 1 + \alpha \cos(k \phi)\,,
\end{equation}
where $k$ is the dimension of the elementary constituent of the heavy operator and $\alpha$ controls the number of traces, $p/N^2$. The results of the holographic analysis of the latter LLM solution, which was  performed in \cite{Skenderis:2007yb} at finite $\alpha$, indicate, however, that this conjecture does not pass a basic consistency check, which has been previously used and tested in the AdS$_3$ context in \cite{Ganchev:2023nef}. It was indeed found in \cite{Skenderis:2007yb} that the single-particle neutral CPO of dimension $\Delta=4$ has a non-vanishing expectation value in the LLM geometry associated with the monochromatic droplet \eqref{eq:monodrop}. This leads to an inconsistency for $k=2$ because, as we will explain in more detail in Section~\ref{sec:heavysearch}, if the LLM geometry were dual to the monochromatic multi-trace, it would imply that the extremal 3-point correlator containing the dimension 4 CPO and two of the dimension 2 constituents of the multi-trace does not vanish, contradicting the general rule that holographic extremal correlators should vanish. This argument suggests that the multi-trace dual to the LLM geometry \eqref{eq:monodrop} for $k=2$ contains also single-trace constituents of dimension 4 and, conversely, that to construct the geometric duals to monochromatic multi-traces one should add further harmonics to the profile \eqref{eq:monodrop} at higher orders in $\alpha$. Moreover, even if we knew the proper generalisation of \eqref{eq:monodrop}, it would probably be unfeasible to compute the integrals necessary to express the LLM solution in closed form for finite $\alpha$. 

To get around these difficulties, we have sought inspiration in the AdS$_3$ context and in particular in the constructions based on a consistent truncation of the 6D supergravity on AdS$_3\times S^3$ down to a 3D gauged supergravity with asymptotically AdS$_3$ solutions \cite{Mayerson:2020tcl,Houppe:2020oqp,Ganchev:2021pgs,Ganchev:2021iwy,Ganchev:2021ewa,Ganchev:2022exf,Ganchev:2023nef}. Working in a truncation limits the degrees of freedom to the lowest harmonics on $S^3$ but it has the advantage -- crucial for our problem -- to restrict the ambiguities one encounters in constructing a solution of the supergravity equations that extends a given linearised solution. The intuitive idea is that the non-linear solution that fits the truncation and reduces at linear order to the perturbation dual to a given CPO describes precisely the monochromatic multi-trace constructed out of the CPO. We will show in the following that a 5D consistent truncation can effectively be used also to construct geometries dual to $\mathcal{N}=4$ SYM heavy states. The AdS$_5$ context is even simpler than the AdS$_3$ one because, while in this latter case the truncation preserves degrees of freedom from both the gravity and the tensor multiplets and this leaves some ambiguity in the solution of the 3D supersymmetric equations, the supergravity multiplet exhausts all the content of the AdS$_5\times S^5$ theory and, hence, the truncation we will use allows no ambiguity when solving the equations. 

After restating in more quantitative terms our problem and goals in Section~\ref{sec:prob}, Section~\ref{sec:trunc} describes the 5D truncation in which we will construct our solution: it is a subtruncation of the $SO(6)$ supergravity found in \cite{Cvetic:2000nc} that is tailored to have the symmetries implied by the monochromatic multi-trace constructed out of the lowest CPO of $\mathcal{N}=4$ SYM (of dimension $\Delta=2$). In Section~\ref{sec:pert} we start from the linearised solution of the equations of motion dual to this CPO and we construct in perturbation theory its non-linear extension. In CFT terms, the perturbation parameter controls the ratio between the number of traces in the heavy state and the central charge, $p/N^2$.  We find that, at all perturbative orders, there is a unique solution that is regular and normalisable or, more precisely, has the asymptotic limit appropriate to describe a state of the CFT without any relevant or irrelevant source. The structure of the perturbative solution is closely analogous to that found in the AdS$_3$ context. For larger values of the perturbation parameter, however, we expect a qualitative difference. In AdS$_3$ the infinite-dimensional 2D chiral algebra implies a ``stringy exclusion principle" \cite{Lerche:1989uy,Maldacena:1998bw} which puts an upper bound on the admissible values of $p$; on the gravity side, violating this bound causes the emergence of closed time-like curves. No such bound exist in AdS$_5$, where the superconformal algebra is finite-dimensional, and thus we expect to find regular solutions for arbitrary, unbounded values of $p/N^2$. To verify this expectation it is necessary to go beyond perturbation theory, which we do in Section~\ref{sec:susy} by taking advantage of the constraints implied by supersymmetry. We construct an exact solution expressed in terms of one function of one variable which, however, we can only define implicitly as the solution of a transcendental equation. To gain some intuition on the solution, we study some of its limits, both as a function of $p/N^2$ and of the radial coordinate, in Section~\ref{sec:lim}. All evidence supports the existence of a proper geometry for arbitrarily large $p/N^2$. In Section~\ref{sec:holo} we perform a preliminary holographic study of our solution and compute its energy and R-charge and the expectation values of the CPO's of dimension 2 and 4 -- for this purpose we apply the gauge-invariant formalism in \cite{Skenderis:2007yb}. The main outcome of this study is that it provides a first non-trivial check that our solution is dual to the monochromatic multi-trace operator we are after. Indeed, contrary to the LLM geometry with profile \eqref{eq:monodrop}, in our solution the expectation value of the neutral CPO of dimension 4 vanishes, as dictated by the vanishing of the extremal 3-point coupling. The concluding Section~\ref{sec:conc} contains a summary of our main results and some comments on possible developments, including the significance of our geometry for the understanding of the black hole microstates. Some details of the computations are provided in the Appendix: the equations of motion of the 5D truncation are listed in Appendix~\ref{app:eom} and the holographic derivation of the R-charge and the CPO expectation values are described in Appendix~\ref{app:holo}.

\section{The problem}
\label{sec:prob}
In this section we provide some background material that helps to set the problem. We first review the linear perturbation of AdS$_5\times S^5$ \cite{Kim:1985ez} that is dual to a single-trace chiral primary operator of $\mathcal{N}=4$ SYM and we recall \cite{Grant:2005qc} that this solution can be mapped to the LLM geometry \cite{Lin:2004nb} associated with a ripple deformation of the circular droplet, at first order in the deformation. We explain why it is not obvious to identify the non-linear completion of the LLM solution that is dual to the simplest multi-trace operator formed by multiple copies of identical single-trace constituents.

\subsection{Light 1/2 BPS states and linearised deformations}
In the $\mathcal{N}=4$ $SU(N)$ gauge theory, the 1/2 BPS single-trace operators of dimension $\Delta=k$ sit in the $[0,k,0]$ R-symmetry multiplet of 
\begin{equation}\label{eq:singletrace}
\mathcal{O}_k = \mathrm{Tr} Z^k\,,
\end{equation}
where $Z=\Phi^1+ i \Phi^2$ is a complex combination of the adjoint scalars $\Phi^I$ ($I=1,\ldots,6$). The gravity duals of these operators\footnote{More precisely, the CFT expression \eqref{eq:singletrace} is only correct at leading order in $1/N$: for $k\ge 4$ $\mathcal{O}_k$ should include appropriate multi-trace corrections  in order to be dual to a single-particle excitation in gravity \cite{Aprile:2020uxk}. These corrections guarantee the orthogonality between single and multi-particle operators. In the following the operators $\mathcal{O}_k$ will always denote the properly defined single-particle operators, even if we will use single-trace and single-particle interchangeably.} are the linearised deformations of AdS$_5\times S^5$ found in \cite{Kim:1985ez}: if we denote by $\bar g_{\mu\nu}$ ($\mu,\nu=0,\dots,4$) and $\bar g_{\alpha\beta}$ ($\alpha,\beta=1,\dots,5$) the undeformed AdS$_5$ and $S^5$ metrics and by $h_{\mu\nu}$ and $h_{\alpha\beta}$ the respective deformations, one has\footnote{The awkward normalisation is chosen for later convenience.}
\begin{equation}\label{eq:Kimetal}
h_{\mu\nu} = -\epsilon\, \frac{3}{4(k+1)}\,\left[2\nabla_\mu\nabla_\nu b_k-k(k-1) \bar g_{\mu\nu} \,b_k \right] Y_k\quad,\quad h_{\alpha\beta}=-\epsilon \,\frac{3}{4}k\, \bar g_{\alpha\beta} \,b_k\, Y_k\,;
\end{equation}
here $\epsilon$ is the perturbative parameter, in which we expand at first order in this analysis, $\nabla_\mu$ is the covariant derivative with respect to the unperturbed AdS$_5$ metric, and $b_k$ and $Y_k$ are eigenfunctions of the AdS$_5$ and $S^5$ laplacians
\begin{equation}
\Box_{AdS_5} b_k = k(k-4) b_k\quad ,\quad \Box_{S^5} Y_k = -k(k+4) Y_k\,.
\end{equation}
If we introduce the following parametrisations for AdS$_5$ and $S^5$:
\begin{equation}\label{eq:AdS5andS5}
d{\bar s}_{AdS_5}=d\rho^2 -\cosh^2\!\rho \,d\tau^2+\sinh^2\!\rho\, d\Omega_3^2 \quad,\quad d{\bar s}_{S^5}=d\theta^2 +\cos^2 \theta \,d{\tilde \phi}^2+\sin^2\theta \,d{\tilde \Omega}_3^2\,,
\end{equation}
then we can take
\begin{equation}\label{eq:bandY}
b_k = \cosh^{-k}\!\rho\,e^{i k \tau}\quad,\quad Y_k = \cos^k\theta\,e^{i k \tilde \phi}\,.
\end{equation}
We see that the solution only depends on $\phi\equiv \tilde\phi+\tau$ and its isometry group is $SO(4)\times SO(4) \times \mathbb{R}$, corresponding to rotations on the two $S^3$'s, $\Omega_3$ and $\tilde \Omega_3$, and to $\tau$ translations at fixed $\phi$. Note that with the choice \eqref{eq:bandY} the metric perturbation in \eqref{eq:Kimetal} is complex, but when we will consider the non-linear extension of the solution, we will always implicitly assume to take the real part of the linearised solution. With this understanding, the solution acquires a further $\mathbb{Z}_2$ symmetry that acts on the coordinates introduced above as $(\tau,\phi)\to -(\tau,\phi)$; this extra symmetry will play a role in the following.

The solution also requires a deformation of the RR 4-form, $\delta C_4$:
\begin{equation}
\delta C_4= \frac{3}{8}\,\epsilon \,\left[b_k\,{\bar \star}_{S^5} dY_k - Y_k\,{\bar \star}_{AdS_5} db_k \right]\,,
\end{equation}
with ${\bar \star}_{AdS_5}$, ${\bar \star}_{S^5}$ the Hodge duals defined with respect to the unperturbed metrics, though we will mostly focus on the metric deformation in the following.

The most general half-BPS solution of type IIB supergravity with isometry $SO(4)\times SO(4) \times \mathbb{R}$ was found in \cite{Lin:2004nb}, and it is uniquely specified by a droplet configuration in a two-dimensional plane. Hence the linearised solution described here must be a limit of some LLM solution. One can indeed verify \cite{Grant:2005qc} that the LLM geometry associated with a deformed disk droplet, whose boundary is the wavy curve
\begin{equation}\label{eq:dropletone}
r(\phi)=1-\frac{\epsilon}{2} \cos(k \phi)\,,
\end{equation}
is diffeomorphic to the geometry in \eqref{eq:Kimetal} at first order in $\epsilon$.

\subsection{The search for heavy states}
\label{sec:heavysearch}
We are interested in 1/2 BPS heavy states in the classical supergravity limit, defined by taking 
\begin{equation}\label{eq:heavylimit}
c = \frac{N^2}{4} \gg 1\quad ,\quad \Delta\gg 1 \quad, \quad \frac{\Delta}{N^2} \quad \mathrm{fixed}\,.
\end{equation}
On the CFT side one can construct such states by taking multi-traces of the light CPO's introduced above
\begin{equation}\label{eq:heavyopkp}
\mathcal{O}_{k,p} = (\mathrm{Tr} Z^k)^p\,,
\end{equation}
with $k$ finite and $p\sim N^2$. We are thus working in a double-scaling limit of large charge $p$ and large $N$, with fixed $p/N^2$. Note that for $D=4$ superconformal theories there is no upper bound on $p$: this is an important qualitative difference with respect to the $D=2$ theories, where the infinite-dimensional superconformal algebra implies a ``stringy exclusion bound" on $p$ of order $c$ \cite{Maldacena:1998bw}. Moreover, since we are taking $N\gg k$, we can discard the trace identities that would otherwise relate operators with different values of $k$ and $N$.

The goal of this article is to find the supergravity description of the states $\mathcal{O}_{k,p}$ in the classical heavy limit \eqref{eq:heavylimit}. As explained in \cite{Skenderis:2006ah,Skenderis:2007yb}, the states that admit a classical supergravity dual are actually coherent state superpositions of the states $\mathcal{O}_{k,p}$, centered over some average value of $p$ but with a non-vanishing spread over a finite range of $p$'s. One should thus consider states of the form
\begin{equation}\label{eq:coherent}
\mathcal{O}_{k,\epsilon} = \sum_{p=0}^\infty c_p(k,\epsilon)\,\mathcal{O}_{k,p} \,,
\end{equation}
where $\epsilon$ is the continuous parameter characterising the coherent state, that should be identified with the $\epsilon$ appearing in \eqref{eq:Kimetal}, and the precise form of the coefficients $c_p(k,\epsilon)$ for the heavy state we are after is currently unknown to us (for the LLM state with profile \eqref{eq:dropletone}, see eq. (6.30) of \cite{Skenderis:2007yb}). In the classical heavy limit \eqref{eq:heavylimit}, the above sum over $p$ peaks over a central value $\bar p$ determined by $\epsilon$, and one expects that $\bar p/N^2$ increases with $\epsilon$. For most purposes in this article, like the computation of simple observables that do not mix different values of $p$ (as the energy or the R-charge) we can ignore this complication and assume the simple form for the heavy operators given in \eqref{eq:heavyopkp}, with the understanding that it represents only the leading term in a coherent state superposition. More complicated observables are sensitive to the $p$-spread, and we will also consider some such observables in the following.  

As a starting point for the construction of the bulk dual of $\mathcal{O}_{k,\epsilon}$, one could consider the limit where $\bar p/N^2$ is made small and formally taken to zero; then the heavy operator $\mathcal{O}_{k,\epsilon}$ effectively becomes light and it should approximate the single-trace CPO $\mathcal{O}_{k}$. That this light limit is continuous is not obvious, but this is what was found in the study of asymptotically AdS$_3$ heavy states -- see for example \cite{Ceplak:2021wzz}, where the continuity of the light limit was used to deduce LLLL correlators from HHLL ones -- and we will assume the same is true for AdS$_5$. Under this hypothesis, the geometry dual to $\mathcal{O}_{k,\epsilon}$ should reduce to the deformation of AdS$_5\times S^5$ given \eqref{eq:Kimetal} in the limit $\bar p/N^2\to 0$, which, as explained above, should be equivalent to sending $\epsilon\to 0$. Our problem then reduces to finding a non-linear completion of the geometry in \eqref{eq:Kimetal} that solves the supergravity equations exactly in $\epsilon$. With only this input the solution is highly non-unique, since when one solves iteratively the supergravity equations order by order in $\epsilon$ starting from \eqref{eq:Kimetal}, one has the freedom to add homogeneous solutions that are not completely fixed even by requiring the regularity of the geometry. The same problem is manifest in the LLM picture, where there are clearly infinitely many choices of droplets that reduce to \eqref{eq:dropletone} at linear order in $\epsilon$. 

One could nevertheless invoke simplicity and argue that the droplet in \eqref{eq:dropletone} gives the correct bulk dual to $\mathcal{O}_{k,\epsilon}$ even for finite $\epsilon$. With this assumption, one can derive the relation between $\epsilon$ and $\bar p/N^2$ by matching the R-charge of $\mathcal{O}_{k,\epsilon}$, $k\,\bar p$, with the one extracted from the asymptotic limit of the LLM geometry defined by the droplet \eqref{eq:dropletone}: this computation was done in \cite{Skenderis:2007yb} with the result that $\epsilon^2 \sim \frac{\bar p}{k\,N^2}$. This fact raises a puzzle: as we remarked above, on the CFT there is no upper bound on $p$, and thus, according to the previous identification between $\bar p$ and $\epsilon$, there should be no upper bound on $\epsilon$ either; on the other hand it is clear from \eqref{eq:dropletone} that when $\epsilon\ge 2$ the droplet degenerates and hence even the bulk dual should develop some pathology. This is hard to verify explicitly because the integrals needed to write down the LLM metric are difficult to compute in closed form for finite $\epsilon$ for the profile \eqref{eq:dropletone}. 

Further evidence against the holographic identification between \eqref{eq:dropletone} and $\mathcal{O}_{k,\epsilon}$ comes from the computation \cite{Skenderis:2007yb} of the expectation value in the state dual to \eqref{eq:dropletone} of the operator $\mathcal{O}_{4,0}$, the descendant of the CPO $\mathcal{O}_{4}$ with vanishing R-charge. Assuming the form \eqref{eq:coherent} for the dual state, the non-trivial part of this expectation value is captured by the 3-point function $\langle \bar{\mathcal{O}}_2 \mathcal{O}_{4,0} \mathcal{O}_{2}\rangle$, where $\mathcal{O}_2$ ($\bar{\mathcal{O}}_2$) is coming from the ket $|\mathcal{O}_{k,\epsilon}\rangle$ (the bra $\langle \mathcal{O}_{k,\epsilon}|$). When $k=2$ this is an extremal 3-point function and thus the expectation value should vanish. Note that the same general criterion has already been tested for asymptotically AdS$_3$ geometries in \cite{Ganchev:2023nef}, where it has provided a consistent interpretation of the so called ``special locus" solution found in \cite{Ganchev:2021pgs} by working in a consistent 3D truncation. Using holographic renormalisation techniques, the authors of \cite{Skenderis:2007yb} have extracted the expectation value of $\mathcal{O}_{4,0}$ from the LLM geometry with profile \eqref{eq:dropletone} and found a non-vanishing result for any $k$, thus contradicting this general criterion. 

In view of these difficulties, we will follow a different strategy, that has proved to be useful in the AdS$_3$ context for both BPS and non-BPS states \cite{Mayerson:2020tcl,Houppe:2020oqp,Ganchev:2021pgs,Ganchev:2021iwy,Ganchev:2021ewa,Ganchev:2022exf,Ganchev:2023nef}. We will work in an effective 5D theory obtained by truncating the 10D space to the lowest $S^5$ harmonics and we will solve the supergravity equations of the truncated theory exactly in $\epsilon$ with the requirement that the solution reduces to \eqref{eq:Kimetal} at linear order in $\epsilon$. We will see that in this restricted setting there is a unique regular and normalisable solution and we will assume that this solution gives the bulk dual of the heavy state $\mathcal{O}_{k,\epsilon}$ for $k=2$.  We will provide a non-trivial test of this assumption by showing that the expectation values of $\mathcal{O}_4$ and its uncharged descendant $\mathcal{O}_{4,0}$, extracted from our geometry with the holographic tools of \cite{Skenderis:2007yb}, are zero, as required by the vanishing of the extremal 3-point functions. 

One obvious limitation of this approach is that the truncated theory only captures the lowest KK harmonics and thus it can only be used to describe the multi-trace operator $\mathcal{O}_{2,\epsilon}$ built with multiple copies of the CPO $\mathcal{O}_{2}$, which is the bottom component of the stress-tensor multiplet of the 5D gauged supergravity.

We will describe the 5D consistent truncation that fits our purposes in the next section.

\section{The 5D truncation}
\label{sec:trunc}
The reduction of type IIB supergravity on $S^5$ leads to the $\mathcal{N}=8$ 5D gauged supergravity that was first found in \cite{Gunaydin:1984qu,Gunaydin:1985cu}. This theory contains a fairly large number of scalars, 42, together with 15 gauge fields and 12 two-form gauge potentials. Since the elementary constituents, $\mathcal{O}_2$, of our heavy state sit in the $20'$ scalar representation of $SO(6)$, and this is the only field that is excited at order $\epsilon$, it is natural to look for a smaller effective theory whose scalar content is restricted to the $20'$. It was indeed shown in \cite{Cvetic:2000nc} that such a consistent truncation exists and it contains, apart from the $20'$ scalars, also the 15 gauge fields and, of course, the 5D metric. Moreover \cite{Cvetic:2000nc} also provides the 10D uplift of the 5D theory, where only the 10D metric and the self-dual five-form are excited. Thus this consistent truncation appears to contain precisely the field content we need. We can actually further simplify the truncation by using the $SO(4)\times SO(4) \times \mathbb{R}\times \mathbb{Z}_2$ symmetry that we have found in the linear solution and that should be preserved at all order in $\epsilon$. Of course, imposing a symmetry guarantees the consistency of this further truncation. 

We will now describe the field content of the supergravity theory we will be working with. We will for the moment concentrate on the 10D metric, but we will also write down the five-form when discussing the supersymmetry. The general form of the 10D metric for the truncation of \cite{Cvetic:2000nc} is
\begin{subequations}\label{eq:truncgen}
\begin{equation}
ds^2_{10}=\Delta^{1/2} ds^2_5 +\Delta^{-1/2} \,T_{ij}^{-1} D\mu^i D\mu^j\,,
\end{equation}
where
\begin{equation}
\Delta = T_{ij}\, \mu^i \mu^j \quad,\quad D\mu^i = d\mu^i + A^{ij} \mu^j\qquad (i,j=1,\ldots,6)\,,
\end{equation}
\end{subequations}
$ds^2_5$ is the asymptotically AdS$_5$ Einstein metric of the 5D base space, the coordinates $\mu^i$, subject to the constraint $\mu^i \mu^i=1$, parametrise the $S^5$, $T_{ij}=T_{ij}$ is a symmetric unimodular ($\mathrm{det} \,T=1$) matrix of zero-forms in 5D, that parametrises the $20'$ scalars, $A^{ij}=-A^{ji}$ are the 15 gauge fields, given by one-forms on the 5D space. The $SO(4)\times SO(4) \times \mathbb{R}\times \mathbb{Z}_2$ symmetry described after \eqref{eq:bandY} restricts the form of these geometric data to the following ansatz:
\begin{subequations}
\begin{equation}\label{eq:5DEinst}
ds^2_5 = \Omega_0^2\left[\frac{d\xi^2}{(1-\xi^2)^2}+\frac{\xi^2}{1-\xi^2} \,d\Omega_3^2\right]-\frac{\Omega_1^2}{1-\xi^2} \,d\tau^2 \,,
\end{equation}
\begin{equation}
T = \mathrm{diag} \left[e^{\lambda-2\mu},e^{-\lambda-2\mu},e^{\mu},e^{\mu},e^{\mu},e^{\mu}\right] \,,
\end{equation}
\begin{equation}\label{eq:gaugeA}
A^{12}=-A^{21}=(1+\Phi)\,d\tau\quad,\quad A^{ij}=0 \quad \mathrm{for}\quad i,j\not=1,2\,,
\end{equation}
\end{subequations}
where all the five unknowns, $\Omega_0$, $\Omega_1$, $\lambda$, $\mu$, $\Phi$, are only functions of the radial coordinate $\xi$; we have traded the coordinate $\rho$ used in \eqref{eq:AdS5andS5} for $\xi = \tanh \rho$ in such a way that asymptotic infinity $\rho\to \infty$ is mapped to the finite point $\xi=1$; we have also fixed the reparametrization invariance related to the radial coordinate by choosing a particular ratio between the $(\xi\xi)$ and the $\Omega_3$ components of the 5D metric $ds^2_5$. Introducing the following parametrization of $S^5$:
\begin{equation}
\mu^1+i\,\mu^2= \cos\theta\,e^{i\phi}\quad,\quad \mu^I = \sin\theta\,\hat x^I\quad \mathrm{for}\quad I=3,\ldots,6\quad\mathrm{with}\quad \hat x^I \hat x^I =1\,,
\end{equation}
we can make explicit the form of the 10D metric for our truncation:
\begin{subequations}\label{eq:truncsmall}
\begin{equation}\label{eq:10mettr}
\begin{aligned}
ds^2_{10}&=\Delta^{1/2} ds^2_5+\Delta^{-1/2}\Bigl[ \left(e^{2\mu}(e^\lambda \sin^2\phi+e^{-\lambda} \cos^2\phi)\sin^2\theta+e^{-\mu}\cos^2\theta\right) d\theta^2\\
&+e^{2\mu}(e^\lambda \cos^2\phi+e^{-\lambda} \sin^2\phi)\cos^2\theta\,D\phi^2 - e^{2\mu} \sinh\lambda\, \sin 2\phi\, \sin 2\theta\,d\theta D\phi +e^{-\mu}\sin^2\theta\,d{\tilde \Omega}_3^2\Bigr]\,,
\end{aligned}
\end{equation}
with
\begin{equation}
\Delta=e^{-2\mu} (e^\lambda \cos^2\phi+e^{-\lambda} \sin^2\phi)\cos^2\theta + e^\mu \sin^2\theta\quad,\quad D\phi= d\phi-(1+\Phi)\,d\tau\,.
\end{equation}
\end{subequations}

The AdS$_5\times S^5$ vacuum is obtained by setting $\Omega_0=\Omega_1=1$, $\lambda=\mu=\Phi=0$; note that the presence of the constant $1$ in the gauge field $A^{12}$ in \eqref{eq:gaugeA} is due to the fact that the $S^5$ coordinate defining the vacuum is $\tilde\phi= \phi-\tau$: this is the coordinate that should be used to read off, for example, the R-charge of the state. However working with the coordinate $\phi$ has the advantage of cancelling all the $\tau$ dependence from the geometry, making the $\mathbb{R}$ isometry manifest.

As a first non-trivial test, one should check that the linear perturbation \eqref{eq:Kimetal} fits in the above truncation for $k=2$. This is not manifest in the form \eqref{eq:Kimetal}, where the perturbation contains non-diagonal terms in the AdS$_5$ part, while in the metric \eqref{eq:10mettr} it is the $S^5$ part that is non-diagonal. This apparent discrepancy is resolved by applying to \eqref{eq:Kimetal} the linearised diffeomorphism generated by the vector field $\xi^M$ ($M\equiv\{\mu,\alpha\}$), with
\begin{equation}
(\xi_\mu,\xi_\alpha)=\frac{\epsilon}{4}\,(\nabla_\mu b_2\,Y_2, - b_2\,\nabla_\alpha Y_2)\,,
\end{equation}
which transforms \eqref{eq:Kimetal} (with $k=2$) into
\begin{equation}
h'_{\mu\nu}=\frac{\epsilon}{2}\,{\bar g}_{\mu\nu}\,b_2\,Y_2\quad,\quad h'_{\alpha\beta}=- \frac{3}{2}\,\epsilon\,b_2 \left[\frac{1}{3}\,\nabla_\alpha \nabla_\beta Y_2 +{\bar g}_{\alpha\beta}\,Y_2\right]\,.
\end{equation}
One can now check that the real part of the perturbation $(h'_{\mu\nu}, h'_{\alpha,\beta})$ matches with  \eqref{eq:truncsmall} for
\begin{equation}\label{eq:firstorderCPO}
\Omega_0=\Omega_1=1\quad,\quad \mu=\Phi=0\quad,\quad \lambda = \epsilon\,(1-\xi^2)\,,
\end{equation}
at first order in $\epsilon$. 

We have thus learnt that the lightest CPO $\mathcal{O}_2$ is described within the truncation \eqref{eq:truncsmall} by a small perturbation of the field $\lambda$. Our next task is to complete this linear solution to an exact non-linear solution of the equations of motion, that is dual to a heavy operator that we conjecture to be the multi-trace $\mathcal{O}_{2,\epsilon}$.

\section{Perturbative solution of the equations of motion}
\label{sec:pert}
In this Section we will solve the equations of motion of our truncation perturbatively in the parameter $\epsilon$ and we will show that there is a unique normalisable solution that reduces to \eqref{eq:firstorderCPO} at first order in $\epsilon$.

The equations of motion for the truncation \eqref{eq:truncsmall} can be derived from the lagrangian
\begin{subequations}
\begin{equation}
\mathcal{L}=\sqrt{-g}\left[R-g^{\mu\nu}\left(\frac{1}{2}\,\partial_\mu\lambda\,\partial_\nu\lambda+3\,\partial_\mu\mu\,\partial_\nu\mu+2\,\sinh^2\lambda \,A^{12}_\mu A^{12}_\nu\right)-\frac{1}{4} e^{4\mu} \,g^{\mu\nu}g^{\rho\sigma} \,F^{12}_{\mu\rho}\,F^{12}_{\nu\sigma}-V\right]\,,
\end{equation}
\begin{equation}
V=2\,e^{-4\mu}\,\sinh^2\lambda - 8 \,e^{-\mu}\,\cosh\lambda-4 \,e^{2\mu}\,,
\end{equation}
\end{subequations}
where $g_{\mu\nu}$ stands for the 5D base metric $ds^2_5$ \eqref{eq:5DEinst} and $F^{12}=dA^{12} = \partial_\xi \Phi\,d\xi\wedge d\tau$.

The equations are written in full in the Appendix (see \eqref{eq:fulleqs}). They consist of five second order ordinary differential equations for the five unknowns, $\Psi\equiv (\lambda,\mu,\Phi,\Omega_0,\Omega_1)$, plus one first order constraint, \eqref{eq:firstordercons}, that originates from having gauge-fixed the diffeomorphism invariance of the 5D metric $ds^2_5$. One can check that the constraint is consistent with the remaining equations by taking its $\xi$-derivative, replacing the second derivatives using the other equations and verifying that one obtains an identity. 

The equations are coupled and non-linear and it is thus hard to envisage a method to solve them exactly, in general. In the next Section, we will eventually construct an exact solution, though expressed in a somewhat implicit form, by leveraging the constraints implied by supersymmetry. It is however useful to start from a more direct perturbative approach, starting from the first order solution \eqref{eq:firstorderCPO} dual to $\mathcal{O}_2$. We thus assume the following $\epsilon$ expansion for the fields
\begin{equation}\label{eq:pertfields}
\lambda=\sum_{n=0} \lambda^{(2n+1)}\epsilon^{2n+1}\quad,\quad \mu = \sum_{n=1} \mu^{(2n)}\epsilon^{2n}\quad,\quad \Phi= \sum_{n=1} \Phi^{(2n)}\epsilon^{2n}\quad,\quad \Omega_i= 1+ \sum_{n=1}\omega_i^{(2n)}\epsilon^{2n}\quad (i=0,1)\,,
\end{equation}
with the input
\begin{equation}\label{eq:lambda1}
\lambda^{(1)}=1-\xi^2\,.
\end{equation}
The equation for the generic field $\Psi$ at order $\epsilon^n$ for $n>1$  has the form
\begin{equation}\label{eq:orderngen}
\nabla^2 \Psi^{(n)}= S_n[\Psi^{(m)}]\,,
\end{equation}
where $\nabla^2$ is the second order wave operator defined in \eqref{eq:waveops} and the source $S_n[\Psi^{(m)}]$ is a function of the fields and their first derivatives of order $m\le n-1$. The equations can thus be solved iteratively\footnote{The source for the field $\omega_1^{(n)}$ actually depends also on $\omega_0^{(n)}$ and its first derivative at the same order $n$, but this does not represent a problem for the iterative procedure, because one can solve first for $\omega_0^{(n)}$ and then use the solution in the source for $\omega_1^{(n)}$.} order by order in $\epsilon$ by simply inverting the wave operator $\nabla^2$. The existence and uniqueness of the perturbative solution thus rests on the invertibility of the operator $\nabla^2$ restricted to the space of fields with specific boundary conditions. The equations of motion have an obvious invariance under the constant rescaling $\Omega_1\to c\,\Omega_1$, $1+\Phi\to c\,(1+\Phi)$, which corresponds to a rescaling of the time coordinate $\tau\to c\,\tau$: we will fix this arbitrariness by the asymptotic boundary condition $\Omega_1(\xi=1)=1$. Moreover, at every perturbative order one has the freedom to shift $\Phi\to \Phi+c'$, which is equivalent to modify the constant $1$ that we introduced in $A^{12}_\tau$. One sees, however, that the constant in $\Phi^{(n)}$ determines the existence of a normalisable solution for $\lambda^{(n+1)}$ and it turns out, quite non-trivially, that the normalisability and regularity of $\lambda$ requires setting to zero this constant at every perturbative order, in such a way that the boundary condition $\Phi(\xi=1)=0$ remains undeformed for generic $\epsilon$. The requirement that the solution asymptotes to AdS$_5\times S^5$ imposes that $\lambda(\xi=1)=\mu(\xi=1)=0$ and $\Omega_0(\xi=1)=1$. We will ask for the more stringent condition that the asymptotic decay of the fields is normalisable in the AdS/CFT sense, i.e. that their behaviour at infinity is that associated with a VEV (the expectation value of the field) rather than with a source. For example a field of dimension two, like $\lambda$, should decay like $1-\xi^2$ but not like $(1-\xi^2)\log(1-\xi^2)$. And we will require all the fields to be regular in the whole allowed range of $\xi$, $0\le\xi\le1$. One can easily verify that the only field for which there exist regular and normalisable solutions, in the sense explained above, of the homogeneous wave equation, $\nabla^2\,\Psi=0$, is $\lambda$, whose homogeneous solution is the linear order deformation $\lambda^{(1)}=1-\xi^2$. Consider, for example, the field $\mu$, which is dual to the R-symmetry descendant of $\mathcal{O}_2$ with vanishing R-charge, whose associated spherical harmonic is $Y_{2,0}\sim 3\sin^2\theta-2$: the only regular solution of the equation $\nabla^2\,\mu=0$ \eqref{eq:muhom} behaves at infinity like a source, and we will impose the absence of such terms at every perturbative order. The asymmetry between $\lambda$ and $\mu$ is clearly due to the fact that, when reducing from \eqref{eq:truncgen} to \eqref{eq:truncsmall}, we have broken $SO(6)$ down to $SO(4)$. In other words, in the minimal truncation \eqref{eq:truncsmall} the only field\footnote{This is different from the truncation used in the AdS$_3$ context \cite{Mayerson:2020tcl,Houppe:2020oqp} where, for supersymmetric solutions, one could freely choose the VEVs of two fields, dual to operators of dimension one and two. Indeed the 6D supergravity whose reduction gives rise to the AdS$_3$ theory contains both a gravity and several tensor multiplets and the two independent fields that survive the truncation are the lowest dimension members of a tensor multiplet and of the gravity multiplet. The field content of the AdS$_5$ theory is, on the other hand, fully contained in the gravity multiplet. In this sense, the solution we construct here should be thought as the analogue of the ``purely NS superstratum" of \cite{Ganchev:2021iwy,Ganchev:2022exf}, where only the VEV of the dimension two field is turned on.} for which we can freely turn on a VEV is the one dual to $\mathcal{O}_2$; the VEVs of all other fields are determined by the non-linear backreaction in terms of this primary VEV. This observation has the important mathematical consequence that at every order in $\epsilon$ the solution of the equations \eqref{eq:orderngen} is unique.\footnote{Adding an arbitrary homogeneous solution for $\lambda^{(n)}$ just amounts to redefine the parameter $\epsilon$ and we will arbitrarily fix this freedom.} The existence of a solution $\Psi^{(n)}$, fulfilling the regularity and asymptotic conditions, is instead highly non-obvious, a priori. It is computationally quite easy to push the iterative solution of the equations to relatively high perturbative order (we have reached order $\epsilon^{21}$) and we have gathered a convincing empirical evidence that the solution exists for arbitrary $n$ and it has the form of a polynomial in $\xi$.

To summarise, there exists a unique, regular and normalisable perturbative solution of the equations of motion, that reduces to the solution for the CPO $\mathcal{O}_2$, \eqref{eq:lambda1}, at $O(\epsilon)$. At order $\epsilon^n$ the field $\Psi^{(n)}$, with  $\Psi^{(n)}$ any of the fields $\lambda^{(n)}$, $\mu^{(n)}$, $\Phi^{(n)}$, $\omega_i^{(n)}$ defined in \eqref{eq:pertfields}, has the form
\begin{equation}
\Psi^{(n)}=(1-\xi^2)\,{\tilde\Psi}^{(n)}\,,
\end{equation}
where ${\tilde\Psi}^{(n)}$ is a polynomial in $\xi^2$ of degree growing like $2n$. We give these polynomials up to order $\epsilon^5$:
\begin{equation}\label{eq:pert2}
{\tilde\mu}^{(2)}=\frac{1}{6}\quad,\quad {\tilde \Phi}^{(2)}=-\frac{1}{2}\quad,\quad \tilde{\omega}_0^{(2)}=0\quad,\quad {\tilde \omega}_1^{(2)}=-\frac{1}{6}(1-\xi^2)\quad,\quad {\tilde \lambda}^{(3)}=\frac{1}{6}\xi^4\,,
\end{equation}
\begin{equation}\label{eq:pert4}
\begin{aligned}
&{\tilde\mu}^{(4)}=-\frac{1}{72}\left(2-2\,\xi^2-\xi^4\right)\,\,,\,\, {\tilde \Phi}^{(4)}=\frac{1}{24}\left(5-5\,\xi^2-\xi^4\right)\,\,,\,\, \tilde{\omega}_0^{(4)}=-\frac{1}{504}(1-\xi^2)(6+\xi^2)\,,\\
&{\tilde \omega}_1^{(4)}=\frac{1}{504}(1-\xi^2)(21-20\,\xi^2-15\,\xi^4)\,\,,\,\, {\tilde \lambda}^{(5)}=\frac{1}{2520}\,\xi^2\,(4-26\,\xi^2-86\,\xi^4+129\,\xi^6)\,.
\end{aligned}
\end{equation}
It is interesting to compare the above perturbative solution with the LLM geometry with the profile \eqref{eq:dropletone}, which can be constructed explicitly by performing the necessary integrals order by order in $\epsilon$. One can check that the two solutions differ already at order $\epsilon^2$ and that the LLM droplet needed to reproduce \eqref{eq:pert2}-\eqref{eq:pert4} at that order is instead 
\begin{equation}
r(\phi)=1-\frac{\epsilon}{2} \cos(2 \phi) + \frac{3}{16} \,\epsilon^2 \cos(4 \phi) + O(\epsilon^3)\,.
\end{equation}
This suggests that the multi-trace operator formed only by powers of $\mathcal{O}_2$ is dual to an LLM geometry generated by a complicated droplet, containing an infinite series of harmonics, $\cos( k\phi)$, and that, conversely, the simple profile \eqref{eq:dropletone} maps to a multi-trace that involves several CPO's, $\mathcal{O}_k$ with $k\ge 2$.

\section{Supersymmetry and the exact solution}
\label{sec:susy}
The perturbative approach of the previous Section gives persuasive evidence of the existence and uniqueness of a solution with the properties expected for the bulk dual of $\mathcal{O}_{2,\epsilon}$, and provides a constructive way to determine the solution in the limit in which the CFT state reduces to a light multi-trace operator formed by a finite number, $\bar p\ll N^2$, of single-trace constituents; in this regime the geometry describes a deformation around AdS$_5\times S^5$. To study the large charge ($\bar p\gtrsim N^2\gg 1$) limit of the geometry one needs, however, to go beyond the regime of small $\epsilon$. It is natural to expect that to construct an exact solution the simplifications implied by supersymmetry might be crucial. In the perturbative approach, supersymmetry enters only through the first order input, \eqref{eq:lambda1}, which describes the supersymmetric single-trace $\mathcal{O}_2$. If our working assumption that the 1/2-BPS state $\mathcal{O}_{2,\epsilon}$ is described within the minimal truncation \eqref{eq:truncsmall} is correct, the uniqueness of the perturbative solution implies, however, that the geometry we have constructed in the previous section preserves 16 supercharges at all orders in $\epsilon$. In this Section, we will verify this expectation, thus lending further support to our assumption.

We will not attempt here a systematic analysis of the supersymmetry constraints in the truncation \eqref{eq:truncgen}, nor in its minimal version \eqref{eq:truncsmall}, and, to the best of our knowledge, such an analysis is not available in the literature. We follow a somewhat brute-force route, which exploits some result of the Appendix A of \cite{Lin:2004nb}. For this purpose, we quote their notation for the 10D metric, $ds^2_{10}$, and the five-form, $F_5$:
\begin{equation}
ds^2_{10}=ds^2_4 + e^{H+G}\,d\Omega_3^2 + e^{H-G}\,d{\tilde \Omega}_3^2\quad,\quad F_5  = F\wedge d\Omega_3 + {\tilde F}\wedge d{\tilde \Omega}_3\,, 
\end{equation}
where $d\Omega_3^2$ and $d{\tilde \Omega}_3^2$ are the metrics of the $S^3$'s inside the AdS$_5$ and the $S^5$ part of the geometry, $d\Omega_3$ and $d{\tilde \Omega}_3$ are their volume three-forms, $ds^2_4$ is the metric in the four-dimensional subspace spanned by $\xi$, $\tau$, $\theta$ and $\phi$, $H$, $G$ are scalars and $F$, $\tilde F$ are closed two-forms in this 4D subspace. The values of $ds^2_4$ and of $H$, $G$ in our minimal ansatz can be read off from \eqref{eq:truncsmall} and \eqref{eq:5DEinst}, in particular:
\begin{equation}
e^{H+G}=\frac{\xi^2}{1-\xi^2}\,\Delta^{1/2}\,\Omega_0^2\quad,\quad e^{H-G}=\sin^2\theta\,\Delta^{-1/2}\,e^{-\mu}\,.
\end{equation}
The form of $F$ and $\tilde F$ can be extracted from the uplift formulas\footnote{With respect to the conventions of \cite{Cvetic:2000nc}, we have changed the normalisation of $F_5$ by a factor $1/4$, so that $F_5 \to \mathrm{vol}(\mathrm{AdS}_5) + \mathrm{vol}(S^5)$ in the AdS limit. We are choosing our orientation in such a way that 
\begin{equation}
\mathrm{vol}(\mathrm{AdS}_5) = - \frac{\xi^3}{(1-\xi^2)^3}\,d\xi\wedge d\tau\wedge d\Omega_3\quad ,\quad \mathrm{vol}(S^5) =  \sin^3\theta\cos\theta\,d\theta\wedge d\phi\wedge d{\tilde \Omega}_3\,,
\end{equation}
to conform with the convention of \cite{Lin:2004nb}.} in \cite{Cvetic:2000nc}; it is easier to write the corresponding one-form potentials, $B$ and $\tilde B$ (with $F= d B$, ${\tilde F}=d {\tilde B}$) which can be chosen as 
\begin{equation}
\begin{aligned}
B&=\frac{1}{8}\,\frac{\xi^3\,\Omega_0^4}{(1-\xi^2)^2\,\Omega_1}\,\sinh(2\lambda)\,(1+\Phi)\,\sin(2\phi)\cos^2\theta\,d\xi-\frac{1}{8}\,\frac{\xi^3\,\Omega_0^2}{\Omega_1}\,e^{4\mu}\,\partial_\xi\Phi\,\cos^2\theta\,D\phi\\
&+\left(b-\frac{1}{8}\frac{\xi^3\,\Omega_0^2\,\Omega_1}{(1-\xi^2)}\left( \partial_\xi \lambda\,\cos(2\phi) -3\,\partial_\xi\mu \right)\cos^2\theta\right) \!d\tau\quad \mathrm{with}\quad \partial_\xi b=\frac{1}{2}\,\frac{\xi^3\,\Omega_0^4\,\Omega_1}{(1-\xi^2)^3}\,(e^{2\mu}+e^{-\mu}\,\cosh\lambda)\,,\\
{\tilde B}&=-\frac{1}{4}\,\Delta^{-1} \left(e^{-2\mu}\,\sinh\lambda\,\sin(2\phi)\sin^3\theta \cos\theta\, d\theta+e^\mu \sin^4\theta\,D\phi\right)\,.
\end{aligned}
\end{equation}
As a check of the above expressions one can verify the self-duality of $F_5$, which translates into 
\begin{equation}
\tilde F= -e^{-3G}\, \star_4 F\,,
\end{equation}
with $\star_4$ the Hodge dual associated with $ds^2_4$.

It is shown in \cite{Lin:2004nb} (see Eq. (A.48)) that supersymmetry implies
\begin{equation}\label{eq:susymain}
d(e^{-(H+G)})\wedge F + d(e^{-(H-G)})\wedge {\tilde F}=0\,.
\end{equation} 
Note that the l.h.s. of this equation depends non-trivially on the $S^5$ coordinates and, hence, requiring it to vanish identically in $(\theta,\phi)$ imposes strong constraints on our five unknown functions of $\xi$, though it requires some effort and a certain amount of guessing to extract them. Even if there might be a more systematic and efficient way to derive them, it turns out that these constraints are enough to allow for an exact solution of our problem. We find that a set of sufficient conditions to satisfy the supersymmetry requirement \eqref{eq:susymain}, together with the first order constraint \eqref{eq:firstordercons}, are 
\begin{subequations}\label{eq:susycons}
\begin{equation}\label{eq:susyalg}
1+\Phi= e^{-3\mu}\quad,\quad e^{2\mu}(\xi^2\,\Omega_0^2-\Omega_1^2)+1-\xi^2=0 \,,
\end{equation}
\begin{equation}\label{eq:susydiff}
\begin{aligned}
\partial_\xi \lambda=-2\,\frac{\xi\,\Omega_0^2}{(1-\xi^2)\,\Omega_1}\,e^{-2\mu}\,\sinh\lambda\quad,\quad \partial_\xi\mu=\frac{2}{3} \,\frac{\Omega_1}{\xi(1-\xi^2)}\,e^{-2\mu}(\cosh\lambda-e^{3\mu})\,,\\
\partial_\xi \Omega_1=\frac{1}{3}\,\frac{\Omega_1^2}{\xi(1-\xi^2)}\,e^{-2\mu}(\cosh\lambda+2\,e^{3\mu})-\xi^{-1}e^{-4\mu}\,\cosh\lambda-\frac{\xi\,\Omega_1}{1-\xi^2}\,.\qquad
\end{aligned}
\end{equation}
\end{subequations}
The first set of constraints, \eqref{eq:susyalg}, are algebraic and allow to eliminate two of the unknowns, say $\Phi$ and $\Omega_0$; the second set, \eqref{eq:susydiff}, are three first order differential relations for the remaining three unknowns. One can check that \eqref{eq:susyalg} and \eqref{eq:susydiff} imply all the five second order equations of motion, \eqref{eq:eoms1}-\eqref{eq:eoms5}, and thus they completely encapsulate our supergravity problem. We have also checked (up to order $\epsilon^{21}$) that the perturbative solution constructed in the previous section satisfies \eqref{eq:susyalg} and \eqref{eq:susydiff}, thus verifying the expectation that the perturbative solution describes a supersymmetric state. 

To recap, supersymmetry has significantly simplified the problem from the five second order equations \eqref{eq:eoms1}-\eqref{eq:eoms5} to the three first order equations \eqref{eq:susydiff} in the unknowns $\lambda$, $\mu$ and $\Omega_1$. These equations, however, are still coupled and non-linear and some more effort is needed to reduce the system to quadrature. A first useful observation is that the system \eqref{eq:susydiff} has an integral of motion:
\begin{equation}
\mathcal{H}\equiv\frac{\Omega_1^2}{1-\xi^2}\,e^{2\mu}\,\sinh\lambda \quad \mathrm{with}\quad \partial_\xi \mathcal{H}=0\,.
\end{equation}
A priori, the constant value of $\mathcal{H}$ should be left arbitrary and determined at the end by the regularity and normalizability of the solution. We find it more convenient to extrapolate the proper value from the perturbative solution, which suggests to take:
\begin{equation}\label{eq:Hconst}
\mathcal{H}=\sinh\epsilon\,.
\end{equation}
Of course we will be able to confirm this guess a posteriori. With the new input, \eqref{eq:Hconst}, the system further reduces to two coupled first order equations for two unknowns, $\lambda$ and $\mu$:
\begin{equation}
\partial_\xi\lambda=\frac{2\,e^{-3\mu} \sinh\epsilon}{\xi (1-\xi^2)^{\frac{1}{2}}}\,\left(\frac{\sinh\lambda}{\sinh\epsilon}\right)^{\frac{1}{2}}\left(\frac{\sinh\lambda}{\sinh\epsilon}-1\right)\quad,\quad \partial_\xi \mu = \frac{2\,e^{-3\mu}}{3\,\xi (1-\xi^2)^{\frac{1}{2}}}\,\left(\frac{\sinh\lambda}{\sinh\epsilon}\right)^{-\frac{1}{2}}(\cosh\lambda-e^{3\mu})\,.
\end{equation}
One can derive $\mu$ from the first equation and substitute it in the second to arrive at a non-linear second order equation for $\lambda$, which can be recast in the form:
\begin{equation}
\partial_\xi \Xi=0\quad \mathrm{with}\quad \Xi\equiv \xi (1-\xi^2)^{\frac{1}{2}} (\sinh \lambda)^{-\frac{3}{2}}\partial_\xi \lambda + 2\,(\sinh\epsilon)^{\frac{1}{2}}\coth\lambda\,.
\end{equation}
Thus, we finally arrive at a first order equation for $\lambda$: $\Xi=\mathrm{const.}$, which can be solved by quadrature. As before, the value of the constant could be determined a posteriori, but it is simpler to guess it from the perturbative solution, obtaining:
\begin{equation}
\Xi=2\,(\sinh\epsilon)^{-\frac{1}{2}}\cosh\epsilon\,.
\end{equation}
Solving the first order equation requires the integral
\begin{equation}\label{eq:lambdaimpl}
\begin{aligned}
&\int_0^\lambda \frac{dx}{(\sinh x)^{\frac{1}{2}}(\tanh\epsilon\cosh x-\sinh x)} =\frac{2}{3}\,\coth^2\!\epsilon \,(\sinh\lambda)^{\frac{3}{2}} F_1\left(\frac{3}{4},\frac{1}{2},1,\frac{7}{4};-\sinh^2\lambda,\frac{\sinh^2\lambda}{\sinh^2\epsilon}\right)\\
&\qquad +\frac{\cosh\epsilon}{(\sinh\epsilon)^{\frac{1}{2}}}\left(\arctan \left[\left(\frac{\sinh\lambda}{\sinh\epsilon}\right)^{\frac{1}{2}}\right]+ \arctanh \left[\left(\frac{\sinh\lambda}{\sinh\epsilon}\right)^{\frac{1}{2}}\right]\right)\,, 
\end{aligned}
\end{equation}
where $F_1$ is the Appell function of the first kind.
The function $\lambda$ (with the asymptotic boundary condition $\lambda(\xi=1)=0$), is then found by inverting the following equation: 
\begin{equation}\label{eq:lambdasol}
\begin{aligned}
\arctanh\,[(1-\xi^2)^{\frac{1}{2}}]&=\frac{1}{3} \cosh\epsilon \,\left(\frac{\sinh\lambda}{\sinh\epsilon}\right)^{\frac{3}{2}}F_1\left(\frac{3}{4},\frac{1}{2},1,\frac{7}{4};-\sinh^2\lambda,\frac{\sinh^2\lambda}{\sinh^2\epsilon}\right)\\
&+\frac{1}{2}\arctan \left[\left(\frac{\sinh\lambda}{\sinh\epsilon}\right)^{\frac{1}{2}}\right]+\frac{1}{2} \arctanh \left[\left(\frac{\sinh\lambda}{\sinh\epsilon}\right)^{\frac{1}{2}}\right]\,.
\end{aligned}
\end{equation}

\subsection{Summary of the solution}
As the formulas needed to reconstruct the solution are spread throughout this section, we collect them below. Let $\lambda$ be the function implicitly defined in \eqref{eq:lambdasol}. The remaining geometric functions are
\begin{subequations}\label{eq:exactmuetal}
\begin{align}
\mu=\frac{1}{3}\,\log \left(\frac{\sinh\epsilon-\sinh\lambda}{\sinh(\epsilon-\lambda)}\right)\quad&,\quad \Phi=\frac{\sinh(\epsilon-\lambda)}{\sinh\epsilon-\sinh\lambda}-1\,,\\
\Omega_0=\frac{(1-\xi^2)^{\frac{1}{2}}}{\xi}\,\left(\frac{\sinh\epsilon}{\sinh\lambda}-1\right)^{\frac{1}{2}} \left(\frac{\sinh(\epsilon-\lambda)}{\sinh\epsilon-\sinh\lambda}\right)^{\frac{1}{3}}\,\,&,\,\, \Omega_1=(1-\xi^2)^{\frac{1}{2}} \left(\frac{\sinh\epsilon}{\sinh\lambda}\right)^{\frac{1}{2}} \left(\frac{\sinh(\epsilon-\lambda)}{\sinh\epsilon-\sinh\lambda}\right)^{\frac{1}{3}}\,.
\end{align}
\end{subequations}

\section{Limits of the exact solution}
\label{sec:lim}

Since we cannot write $\lambda$ explicitly, the best we can do to analyse the geometry is to consider various limits where $\lambda$ admits an expansion in terms of elementary functions. This will allow us to perform some consistency and regularity checks on the exact solution we have just constructed. We will use the asymptotic expansion of the geometry that we derive in this Section to carry out a holographic analysis of the solution in the next Section.   

The Appell function in \eqref{eq:lambdasol} is a generalisation of the ordinary hypergeometric and it can be expanded in terms of ${}_2 F_1$ functions when one of its last two arguments, $x$ or $y$, goes to zero: 
\begin{equation}\label{eq:Appellxto0}
\begin{aligned}
F_1(a,b_1,b_2,c;x,y)&=\sum_{m=0}^\infty \frac{(a)_m\,(b_1)_m}{(c)_m}\, {}_2F_1(a+m,b_2,c+m;y)\,\frac{x^m}{m!}\\
&=\sum_{n=0}^\infty \frac{(a)_n\,(b_2)_n}{(c)_n}\, {}_2F_1(a+n,b_1,c+n;x)\,\frac{y^n}{n!}\,,
\end{aligned}
\end{equation}
with $(a)_m$ the rising Pochhammer symbol.  Moreover, for the values of  $a$, $b_2$ and $c$ appearing in \eqref{eq:lambdasol} the hypergeometric functions in the first line of \eqref{eq:Appellxto0} reduce to elementary functions:
\begin{equation}\label{eq:2F1x0}
{}_2F_1\left(\frac{3}{4}+m,1,\frac{7}{4}+m;y\right)=\frac{4m+3}{y^m}\left[\frac{\arctanh(y^{\frac{1}{4}})-\arctan(y^{\frac{1}{4}})}{2\,y^{\frac{3}{4}}}-\sum_{p=0}^{m-1} \frac{y^p}{4p+3}\ \right]\,.
\end{equation}
This observation makes some limits of the solution, as a function of either $\epsilon$ or $\xi$, accessible to an explicit and analytical analysis. The variables that appear in the equation for $\lambda$ \eqref{eq:lambdasol} are
\begin{equation}
x= -\sinh^2\lambda\quad,\quad y =\frac{\sinh^2\lambda}{\sinh^2\epsilon}\,,
\end{equation}
and their values control the different regimes. The easiest cases are when $x$ or $y$, or both, go to zero and, hence, one can use the expansions \eqref{eq:Appellxto0} directly. This happens when $\epsilon\to 0$, which is the perturbative regime of Section~\ref{sec:pert}, or when $\xi\to 1$, corresponding to the asymptotic region of the geometry. One can also access the regimes where $x$ or $y$ tend to the other two singular points of the hypergeometric function, which are $1$ and $\infty$. In these situations one can first take the limits of the ${}_2F_1$ functions using the well-known hypergeometric connection formulas and then perform the sums over $m$ or $n$ in \eqref{eq:Appellxto0}. This technique allows us to study the $\xi\to 0$ and $\epsilon\to \infty$ limits. 
 
\subsection{$\epsilon \to 0$}
Consider first the perturbative regime of small $\epsilon$. We know that when $\epsilon\to 0$ $\lambda=O(\epsilon)$ and thus $x\to 0$; one can then use the first line of \eqref{eq:Appellxto0} together with \eqref{eq:2F1x0} to expand the r.h.s. of \eqref{eq:lambdaimpl} to any order in $\epsilon$ and reconstruct the perturbative expansion of $\lambda$. This provides a check of the consistency of the exact solution with the one derived by the brute-force iterative solution of the equations of motion, given in \eqref{eq:pert2}, \eqref{eq:pert4} for the lowest orders. One can also verify that the perturbative solution satisfies the relations \eqref{eq:exactmuetal} for the geometric functions other than $\lambda$.

\subsection{$\xi \to 1$}
The asymptotic expansion of the geometry near the AdS boundary is derived by taking $\xi\to 1$ keeping $\epsilon$ finite. This limit is even simpler to compute, because one expects $\lambda = O(1-\xi^2)$ and thus both the variables $x$ and $y$ are going to zero. Apart from confirming that the geometry has a well-behaved asymptotic limit, this regime is physically interesting because it contains informations on the conserved charges and other holographic VEV's that will be computed in the next Section. After expanding the Appell function, the $\lambda$ equation \eqref{eq:lambdasol} can be solved; this, together with \eqref{eq:exactmuetal}, yields:
\begin{equation}\label{eq:asymptotic}
\begin{aligned}
\lambda&=\sinh\epsilon\,(1-\xi^2)-\frac{4}{3}\,\sinh\epsilon\,\sinh^2\!\left(\frac{\epsilon}{2}\right)(1-\xi^2)^2+O((1-\xi^2)^3)\,,\\
\mu&=\frac{2}{3}\,\sinh^2\!\left(\frac{\epsilon}{2}\right)(1-\xi^2)-\frac{8}{9}\,\sinh^4\!\left(\frac{\epsilon}{2}\right)(1-\xi^2)^2+O((1-\xi^2)^3)\,,\\
\Phi&=-2\,\sinh^2\!\left(\frac{\epsilon}{2}\right)(1-\xi^2)+\frac{14}{3}\,\sinh^4\!\left(\frac{\epsilon}{2}\right)(1-\xi^2)^2+O((1-\xi^2)^3)\,,\\
\Omega_0&=1+\frac{1}{6}\,(3\cosh\epsilon-1)\sinh^2\!\left(\frac{\epsilon}{2}\right)(1-\xi^2)^2+O((1-\xi^2)^3)\,,\\
\Omega_1&=1+\frac{1}{6}\,(3\cosh\epsilon-5)\sinh^2\!\left(\frac{\epsilon}{2}\right)(1-\xi^2)^2+O((1-\xi^2)^3)\,,
\end{aligned}
\end{equation}
where we have kept the terms in the $\xi\to 1$ expansion that are needed to extract the VEV's of operators up to dimension 4.

One observes that the geometry has the allowed asymptotic limit -- i.e. the large distance behaviour expected for a non-trivial state in a CFT undeformed by external sources -- for any finite value of $\epsilon$, supporting the belief that there is no upper bound on $\epsilon$. 

\subsection{$\xi \to 0$}

The absence of pathologies in the asymptotic region for any $\epsilon$ does not exclude, of course, the possibility of a singularity appearing somewhere in the bulk for sufficiently large values of $\epsilon$. It is therefore interesting to have analytical control over the behaviour of the geometry around the center of space, $\xi\approx 0$, away from the perturbative regime of small $\epsilon$. The perturbative computation indicates that $\lambda^{(n)}$ vanishes at $\xi=0$ for $n>1$ -- we have checked that this true up to $n=21$ -- and thus suggests the ansatz
\begin{equation}\label{eq:xi0ansatz}
\lambda = \epsilon - f(\epsilon)\,  \xi^2 + O(\xi^4)\,,
\end{equation}
for some function $f(\epsilon)$ that is determined by \eqref{eq:lambdasol}. With this assumption, $y=1-2 \coth\epsilon \,f(\epsilon)\, \xi^2 + O(\xi^4)$ and thus we are led to study the behaviour of $F_1$ around the singular point $y=1$. This can be done by using the first line of \eqref{eq:Appellxto0}, taking the $y\to 1$ limit of \eqref{eq:2F1x0}
\begin{equation}
{}_2F_1\left(\frac{3}{4}+m,1,\frac{7}{4}+m;y\right) \approx (4m+3) \left[-\frac{\log \xi}{2} +\frac{1}{4} \log\left(\frac{\tanh\epsilon}{2\,f(\epsilon)} \right) - \frac{1}{4} H\!\left(m-\frac{1}{4}\right)\right]\,,
\end{equation}
with $H(x)=\gamma+ \psi(x+1)$ the harmonic number, and then performing the sum over $m$:
\begin{equation}
F_1\left(\frac{3}{4},\frac{1}{2},1,\frac{7}{4};x,y\right)\approx \frac{3}{\cosh\epsilon} \left[-\frac{\log \xi}{2} +\frac{1}{4} \log\left(\frac{\tanh\epsilon}{2\,f(\epsilon)} \right)+ g(\epsilon) \right] \,,
\end{equation}
where
\begin{equation}\label{eq:gdef}
g(\epsilon) = -\frac{\cosh\epsilon}{12} \sum_{m=0}^\infty (-1)^m \frac{\left(\frac{3}{4}\right)_m\,\left(\frac{1}{2}\right)_m}{\left(\frac{7}{4}\right)_m}\, (4m+3)\,H\!\left(m-\frac{1}{4}\right) \,\frac{\sinh^{2m}\epsilon}{m!}
\end{equation}
is a function that we have been unable to evaluate analytically but we believe it is finite for all positive values of $\epsilon$. Using the above limit for $F_1$ in \eqref{eq:lambdasol}, we find that for $\xi\to 0$ the logarithmically divergent terms cancel, confirming the validity of the ansatz \eqref{eq:xi0ansatz}, and the finite terms determine $f(\epsilon)$
\begin{equation}\label{eq:fsol}
f(\epsilon) = \tanh\epsilon \,e^{\frac{\pi}{4} -\frac{3}{2}\log 2+ 2 g(\epsilon)}\,.
\end{equation}
The necessary and sufficient condition for the regularity of the metric at $\xi=0$ is that $f(\epsilon)$ be finite and positive; indeed \eqref{eq:exactmuetal} implies that 
\begin{equation}
\Omega_0^2 = \frac{(\cosh\epsilon)^{\frac{1}{3}}}{\sinh\epsilon} f(\epsilon) + O(\xi^2)\,,
\end{equation}
with all other metric functions finite, and thus if $f(\epsilon)$ were to vanish or become negative the metric would be singular. Though we do not have a closed form expression for $g(\epsilon)$ and hence for $f(\epsilon)$, the identity \eqref{eq:fsol} shows that $f(\epsilon)$ is finite and positive as far as $g(\epsilon)$ stays finite; this is obviously true for $\epsilon \to 0$, where $\eqref{eq:gdef}$ implies $g(\epsilon)\to -\frac{\pi}{8}+\frac{3}{4} \log2$ and thus $f(\epsilon) \to \epsilon$, and also for $\epsilon\to \infty$, where $f(\epsilon)\to e^{\frac{\pi}{2}}$, as we will show in the next subsection. This is a strong indication that no singularity develops near $\xi=0$ for any value of $\epsilon$.

\subsection{$\epsilon \to \infty$}
\label{sec:largeeps}
\begin{figure}[!ht]\centering
	\includegraphics[width=.5\textwidth]{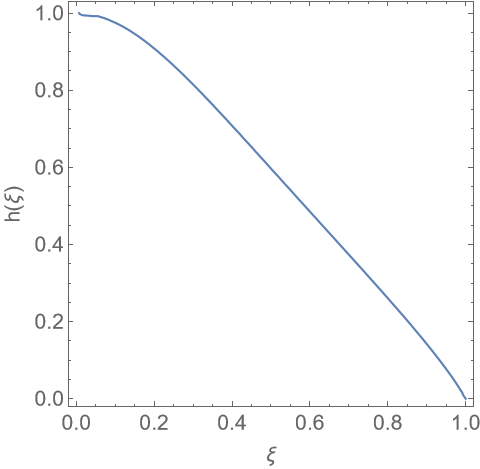}
	\caption{The plot of the function $h(\xi)$ which satisfies the equation \eqref{eq:h}.} 
	\label{fig:Plot}
\end{figure}

It is interesting to examine also the limit of large $\epsilon$ with fixed $\xi$. As we have already remarked, this regime is not present in the AdS$_3$ context and thus it has not yet  been explored on the gravity side. The previous analysis, in particular the $\xi\to 0$ limit, suggests that $\lambda$ diverges when $\epsilon \to \infty$, and thus we should look at the behaviour of $F_1$ for $x\to -\infty$, which is another singular point of the Appell function. We will be able to confirm this working hypothesis a posteriori by verifying its consistency with \eqref{eq:lambdasol}. To derive the $x\to -\infty$ limit one can use the second line of \eqref{eq:Appellxto0} and the known connection formula for the hypergeometric function
\begin{equation}
\begin{aligned}
{}_2F_1\left(\frac{3}{4}+n,\frac{1}{2},\frac{7}{4}+n;x\right)&=\frac{\Gamma\left(\frac{7}{4}+n \right)\Gamma\left(-\frac{1}{4}-n\right)}{\Gamma\left(\frac{1}{2} \right)}(-x)^{-\frac{3}{4}-n}\\
&+\frac{\Gamma\left(\frac{7}{4}+n \right)\Gamma\left(\frac{1}{4}+n\right)}{\Gamma\left(\frac{3}{4}+n\right)\Gamma\left(\frac{5}{4} +n\right)}(-x)^{-\frac{1}{2}} {}_2F_1\left(\frac{1}{2},-\frac{1}{4}-n,\frac{3}{4}-n;\frac{1}{x}\right)\\
&\stackrel{x\to-\infty}{\to}\frac{\Gamma\left(\frac{7}{4}+n \right)\Gamma\left(\frac{1}{4}+n\right)}{\Gamma\left(\frac{3}{4}+n\right)\Gamma\left(\frac{5}{4} +n\right)}(-x)^{-\frac{1}{2}}\,.
\end{aligned}
\end{equation}
Performing the sum over $n$ in \eqref{eq:Appellxto0} gives
\begin{equation}
F_1\left(\frac{3}{4},\frac{1}{2},1,\frac{7}{4};x,y\right)\stackrel{x\to-\infty}{\to}\frac{3}{2} (-x)^{-\frac{1}{2}} \left[\frac{\arctan y^{\frac{1}{4}} +\arctanh y^{\frac{1}{4}} }{y^{\frac{1}{4}}}\right]\,.
\end{equation}
Using this limit in \eqref{eq:lambdasol}, one sees that a consistent ansatz when $\epsilon\to \infty$ is
\begin{equation}\label{eq:largeel}
\lambda \approx \epsilon + \log h(\xi)\,,
\end{equation}
where the function $h(\xi)$ is defined implicitly by
\begin{equation}\label{eq:h}
\arctanh\,[(1-\xi^2)^{\frac{1}{2}}] = \arctanh\,[h(\xi)^{\frac{1}{2}}]+ \arctan\,[h(\xi)^{\frac{1}{2}}]\,.
\end{equation}
In Figure \ref{fig:Plot} we have plotted this implicit function. The limits of $h(\xi)$ for $\xi\to 0$ and $\xi \to 1$ are
\begin{equation}
h(\xi) = 1- e^{\frac{\pi}{2}}\, \xi^2 + O(\xi^4)\quad ,\quad h(\xi) = \frac{1}{4}(1-\xi^2) + O((1-\xi^2)^2)\,.
\end{equation}
Comparing the $\epsilon\to \infty$, $\xi\to 0$ result with \eqref{eq:xi0ansatz} gives the large $\epsilon$ limit of the function $f(\epsilon)$
\begin{equation}
f(\epsilon) \stackrel{\epsilon\to\infty}{\to} e^{\frac{\pi}{2}}\,,
\end{equation}
as we have anticipated in the previous subsection. 

Note that the $\epsilon\to \infty$ result \eqref{eq:largeel} does not lead to an asymptotically AdS geometry. This is not surprising, because taking $\epsilon$ large with fixed $\xi$ excludes the asymptotically AdS region, in which $e^\epsilon\,(1-\xi^2)$ should be taken small, as can be seen by taking $\epsilon$ large in \eqref{eq:asymptotic}. It would be interesting to analyse further the physical meaning of this large $\epsilon$ regime.

\section{Holographic analysis}
\label{sec:holo}
The knowledge of the asymptotic limit of the geometry for arbitrary $\epsilon$, \eqref{eq:asymptotic}, allows us to compute the conserved charges and the expectation values of some CPO's in the dual heavy state; we will dub, for brevity, these expectation values as VEV's, even if it should be kept in mind that they are computed in a non-trivial state and not in the vacuum. These quantities are independent of the coupling constant and thus provide a tool to infer the map between the CFT and the gravity parameters -- in particular the relation between $\bar p/N^2$ and $\epsilon$ -- and also to gather the first non-trivial checks of this holographic map. The basic idea behind these computations is simple and exploits the standard holographic relation between the VEV of an operator of dimension $\Delta$ and the term of order $r^{-\Delta}$ in the large $r$ expansion of the dual supergravity field, where the radial coordinate, $r$, is related to our coordinate $\xi$ as $r^{-2} \sim 1-\xi^2$. However, constructing the precise map between operators and supergravity fields in a coordinate independent way requires a careful analysis that, luckily for us, has been performed in the AdS$_5$ context in \cite{Skenderis:2007yb}, whose results we will heavily borrow in the following. We will report some of the most technical details of the computation in Appendix~\ref{app:holo}.

\subsection{The conserved charges}
We start by computing the energy, $E$, which follows from the holographic formula for the CFT stress tensor $T_{\mu\nu}$ \cite{deHaro:2000wj}. Since the CFT lives on $\mathbb{R}\times S^3$, $T_{\mu\nu}$ has a non-vanishing trace, proportional to the curvature of $S^3$; the trace of the stress-energy tensor gives rise to the non-vanishing vacuum (or Casimir) energy. We will not be interested in this vacuum contribution to the energy, which is independent of the state, and thus we will only focus on the traceless stress-energy tensor, that we denote by ${\tilde T}_{\mu\nu}$. To extract $\tilde T_{\mu\nu}$ from the geometry, one performs the usual Fefferman-Graham expansion of the 5D metric
\begin{equation}
ds^2_5 = \frac{dz^2}{z^2}+\frac{1}{z^2} \,\left(g_{\mu\nu}^{(0)} + z^2 g_{\mu\nu}^{(2)}+z^4 g_{\mu\nu}^{(4)}\right)+O(z^4)\,,
\end{equation}
where $\mu,\nu$ denote here coordinates on the 4D boundary of AdS$_5$ and $g_{\mu\nu}^{(0)}=-d\tau^2+d\Omega_3^2$. Then one has\footnote{In general ${\tilde T}_{\mu\nu}$ also contains terms that depend on $g_{\mu\nu}^{(2)}$. For our solution however $g_{\mu\nu}^{(2)}$ is independent of $\epsilon$: $g_{tt}^{(2)}=g_{ii}^{(2)}=-\frac{1}{2}$, and thus these terms only contribute to the vacuum energy and we omit them.}
\begin{equation}
{\tilde T}_{\mu\nu} = \frac{N^2}{2\pi^2} \left[g_{\mu\nu}^{(4)}-\frac{g_{\mu\nu}^{(0)}}{4} \,g^{(4)\,\lambda}_\lambda\right]\,,
\end{equation}
where indices are raised with $g_{\mu\nu}^{(0)}$. From the asymptotic expansion of the solution, \eqref{eq:asymptotic}, one finds that the Fefferman-Graham coordinate $z$ is related to $\xi$ by
\begin{equation}\label{eq:zxi}
1-\xi^2=z^2 \left[1-\frac{z^2}{2}+\frac{z^4}{48}\left(4+8 \cosh\epsilon-3 \cosh(2\epsilon)\right)\right]+O(z^8)\,,
\end{equation}
and that
\begin{equation}
g_{tt}^{(4)}=\frac{1}{16}(-20+24\cosh\epsilon-5 \cosh(2\epsilon))\quad,\quad g_{ii}^{(4)}=\frac{1}{48}(28-40\cosh\epsilon+15 \cosh(2\epsilon))\quad (i\subset S^3)\,,
\end{equation}
which yields
\begin{equation}
\tilde T_{tt}=\frac{N^2}{2\pi^2} \,\sinh^2\left(\frac{\epsilon}{2}\right)\,.
\end{equation}
Then the energy above the vacuum is obtained by integrating $\tilde T_{tt}$ over $S^3$:
\begin{equation}
E = 2\pi^2\, \tilde T_{tt}=N^2 \sinh^2\left(\frac{\epsilon}{2}\right)\,.
\end{equation}
Our state being supersymmetric, the energy $E$ should equal the R-charge $J$. Verifying this equality is a first consistency check of our construction. 

The R-charge is extracted from the second order term in the Fefferman-Graham expansion of the gauge field $\Phi$:
\begin{equation}
\Phi=\Phi^{(0)}+ z^2\,\Phi^{(2)}+O(z^4)\,.
\end{equation}
Fixing the coefficient of proportionality between $\Phi^{(2)}$ and $J$ requires a bit of work, and we leave the details for the Appendix. The general relation is
\begin{equation}
J=-\frac{N^2}{2}\, \Phi^{(2)}\,,
\end{equation}
which, using \eqref{eq:asymptotic}, gives
\begin{equation}\label{eq:R-charge}
J=N^2 \sinh^2\left(\frac{\epsilon}{2}\right)\,,
\end{equation}
thus confirming the supersymmetric nature of our solution.

The expression for $E=J$ determines the relation between the supergravity parameter $\epsilon$ and the average number of single-trace components of the heavy state, $\bar p$:
\begin{equation}\label{eq:pbar}
\frac{\bar p}{N^2}=\frac{1}{2}\,\sinh^2\!\left(\frac{\epsilon}{2}\right)\,,
\end{equation}
since every single-trace operator $\mathcal{O}_2$ carries $E=J=2$.

\subsection{Expectation values of CPO's}
More stringent checks on the identification of the geometry we have constructed with a state of the form \eqref{eq:coherent} can be obtained by looking at the expectation values in that state of the CPO's $\mathcal{O}_k$ or their $SO(6)$ descendants with R-charge $I$, that we denote by $\mathcal{O}_{k,I}$ (of course $\mathcal{O}_k= \mathcal{O}_{k,k}$ and $\mathcal{\bar{O}}_k= \mathcal{O}_{k,-k}$). Only operators that preserve the $SO(4)$ rotation symmetry acting on the directions $\Phi^i$, with $i=3,\ldots,6$, can have non-trivial expectation values in our state, and thus we can use the descent operator
\begin{equation}
\sum_{i=3}^6 (J^{1i}-i J^{2i})(J^{1i}-i J^{2i})\,,
\end{equation}
with 
\begin{equation}
J^{IJ}\equiv -i \left(\Phi^I \frac{\partial}{\partial \Phi^J} -\Phi^J \frac{\partial}{\partial \Phi^I}\right)\quad I,J = 1,\ldots,6
\end{equation}
the $SO(6)$ generators, to generate $\mathcal{O}_{k,I-2}$ from $\mathcal{O}_{k,I}$, up to a normalisation factor. The normalisation of the single-particle operators $\mathcal{O}_{k,I}$ is the one used in \cite{Skenderis:2007yb}, according to which $\langle \overline{\mathcal{O}}_{k,I} \mathcal{O}_{k,I}\rangle\sim N^2$, while the heavy state $|\mathcal{O}_{2,\epsilon}\rangle$ has unit norm. We will denote the VEV of $\mathcal{O}_{k,I}$  in the heavy state as
\begin{equation}
\langle \mathcal{O}_{k,I} \rangle \equiv  \langle \mathcal{O}_{2,\epsilon} |  \mathcal{O}_{k,I} |  \mathcal{O}_{2,\epsilon} \rangle\,.
\end{equation}

The lightest operator with a non-vanishing VEV is $\mathcal{O}_2$ itself: acting on $\mathcal{O}_{2,p}$, $\mathcal{O}_2$ produces $\mathcal{O}_{2,p+1}$ and since the coherent state $\mathcal{O}_{2,\epsilon}$ contains a sum over $p$, $\mathcal{O}_2$ has a non-trivial expectation value which probes the $p$-dependence of the coefficients $c_p(2,\epsilon)$. The holographic recipe applied to our gravity solution produces indeed a non-vanishing result
\begin{equation}\label{eq:vevo2}
\langle \mathcal{O}_2 \rangle = -\frac{N^2}{2\pi^2} \frac{11\sqrt{2}}{9} \,\sinh\epsilon\,,
\end{equation}
consistently with the CFT expectation. Note however that finding a non-vanishing $\langle \mathcal{O}_2 \rangle$ is not a meaningful test of the non-linear completion of the supergravity solution: any solution that reduces at linear order in $\epsilon$ to the CPO deformation \eqref{eq:Kimetal}, including the LLM geometry with profile \eqref{eq:dropletone}, produces a result of the form of \eqref{eq:vevo2}.  

A more interesting class of observables are the VEVs of the dimension four operators, $\langle \mathcal{O}_{4,I} \rangle$, with $I=0,4$. As already explained at the end of Section~\ref{sec:heavysearch}, the expectation value of $\mathcal{O}_{4,0}$ is proportional to the 3-point coupling $\langle \bar{\mathcal{O}}_2 \mathcal{O}_{4,0} \mathcal{O}_{2}\rangle$, and similarly that of $\mathcal{O}_{4}$ is proportional to  $\langle \bar{\mathcal{O}}_2 \bar{\mathcal{O}}_2 \mathcal{O}_{4}\rangle$ (where now both the $\bar{\mathcal{O}}_2$ operators come from the bra $\langle \mathcal{O}_{k,\epsilon}|$). Both the 3-point couplings above are extremal (the dimension of one of the operators in the correlator equals the sum of the dimensions of the other two), and hence they should vanish. One can explain the vanishing of the extremal couplings by the fact that operators, like $\mathcal{O}_{k,I}$, dual to single-particle supergravity states have to be defined to be orthogonal to all multi-particle states (see for example \cite{Aprile:2020uxk}). This argument then leads to a sharp prediction
\begin{equation}\label{eq:vevo4}
\langle \mathcal{O}_4 \rangle = \langle \mathcal{O}_{4,0} \rangle=0\,.
\end{equation}
We show in Appendix~\ref{app:holo} that our supergravity solution fulfils this prediction and, moreover, the vanishing of $\langle \mathcal{O}_{4,I} \rangle$ is the result of a quite non-trivial cancelation which is sensitive to the non-linear completion of the geometry. As we already mentioned, the LLM profile \eqref{eq:dropletone} is, on the contrary, inconsistent with \eqref{eq:vevo4}. We should mention that a very similar holographic test has been performed in the AdS$_3$ context in \cite{Ganchev:2023nef} (see Section 4.4).

For completeness we should mention that also the VEV of $\mathcal{O}_{2,0}$ is sensitive to the non-linear structure of our solution, being proportional to $\langle \bar{\mathcal{O}}_2 \mathcal{O}_{2,0} \mathcal{O}_{2}\rangle$. The latter 3-point function is not extremal and does not vanish, and indeed the holographic computation produces a non-trivial output: \begin{equation}\label{eq:vevo20}
\langle \mathcal{O}_{2,0} \rangle = \frac{N^2}{2\pi^2} \frac{44\sqrt{2}}{9\sqrt{3}} \,\sinh^2\left(\frac{\epsilon}{2}\right)\,.
\end{equation}
This result is consistent with the interpretation of the dual state as a coherent state of the general form \eqref{eq:coherent}, which leads to the prediction
\begin{equation}\label{eq:vevo20CFT}
\langle \mathcal{O}_{2,0} \rangle_{CFT} =  \frac{\langle \bar{\mathcal{O}}_2 \mathcal{O}_{2,0} \mathcal{O}_{2}\rangle}{\langle \bar{\mathcal{O}}_2  \mathcal{O}_{2}\rangle}\,\bar{p}\,.
\end{equation}
With our normalisation conventions both $\langle \bar{\mathcal{O}}_2 \mathcal{O}_{2,0} \mathcal{O}_{2}\rangle$ and $\langle \bar{\mathcal{O}}_2  \mathcal{O}_{2}\rangle$ are proportional to $N^2$ and thus, using \eqref{eq:pbar}, we see that the CFT prediction \eqref{eq:vevo20CFT} reproduces both the $N$ and $\epsilon$ dependence of the gravity result \eqref{eq:vevo20}; the comparison of the numerical coefficients requires a more careful treatment of the operators normalisations, which we do not perform here. 

Note that deriving a CFT prediction also for the expectation value $\langle \mathcal{O}_2 \rangle$, \eqref{eq:vevo2}, requires the knowledge of the coefficients $c_p(2,\epsilon)$ that appear in the definition of the coherent state \eqref{eq:coherent}. It would be interesting to gather further inputs to help us formulate a guess for such coefficients, but we leave this task for future studies.

\section{Summary and Outlook}
\label{sec:conc}
The main goal of this article has been to construct the supergravity dual of the half-BPS multi-trace operator made by many copies of $\mathcal{O}_2$, the CPO in the stress-tensor multiplet of $\mathcal{N}=4$ SYM, in the large charge regime in which the number of copies is of order $N^2$ in the $N\to \infty$ limit. By working in the 5D consistent truncation singled out by the field content and the symmetries associated with the half-BPS state, we have found an exact solution of the supergravity equations, given in \eqref{eq:exactmuetal}, written in terms of the function $\lambda(\xi)$ defined by the transcendental equation \eqref{eq:lambdasol}. We have shown that the geometry is regular, horizonless and asymptotes AdS$_5$ in a normalisable way for arbitrary values of the R-charge. We have performed the first holographic checks supporting the identification of the state dual to our geometry with the ``monochromatic" multi-trace composed of $\mathcal{O}_2$; in particular we have verified the vanishing of the expectation value of the CPO of dimension 4, as required by the triviality of extremal 3-point correlators. 

Some aspects of our analysis deserve further study. It would be interesting to find a more precise characterisation of the heavy state dual to our geometry. We have conjectured that the state should have the form of the coherent superposition given in \eqref{eq:coherent}, but the expression for the coefficients $c_p$ is still unknown to us. The results of the holographic calculation described in Section~\ref{sec:holo} constrain these coefficients and, hopefully, they will suggest a natural ansatz. Some potentially new insights might also come from the study of the large $\epsilon$ limit of our solution, which we have just touched upon in Section~\ref{sec:largeeps}. This limit describes the very large charge regime, $p\gg N^2$, and it represents a novelty in the microstate literature, since in AdS$_3$ this regime was forbidden by the stringy exclusion principle. 

More broadly, our result opens the possibility to extend to the AdS$_5$ context the pletora of holographic studies performed on AdS$_3$ microstates, with the added bonus of having a much better understood CFT dual. A first direction we intend to pursue is the study of the wave equation describing light supergravity operators in the background provided by our solution. The asymptotic expansion of the solution of this equation encodes the HHLL correlator in the supergravity limit, i.e. large $N$ and large 't Hooft coupling. On the other hand, given the form \eqref{eq:coherent} of our heavy operator, the small $\epsilon$ expansion of the HHLL correlator should yield correlators with multi-trace insertions, of the form $\langle \mathcal{O}_2 \mathcal{O}_2 \mathcal{O}_{2,p} \mathcal{O}_{2,p}\rangle$. Recent developments \cite{Chester:2020dja,Paul:2023rka,Brown:2023why} have allowed the computation of integrated versions of such correlators for arbitrary $N$ and arbitrary 't Hooft coupling and, although our supergravity limit could capture only a small corner of those results, it could still provide a useful\footnote{We thank R. Russo and C. Wen for preliminary useful discussions on this point.} consistency check.

Finally, we would like to comment on the relevance of our findings for understanding the microstates of black holes. To find an asymptotically AdS$_5$ supersymmetric black hole with a finite horizon in supergravity (known as the Gutowski-Reall black \cite{Gutowski:2004yv}) one needs an ensemble of 1/16 BPS states, and our half-BPS solution is obviously quite far from that class of states. The development of the microstate construction for the Strominger-Vafa black hole has, however, shown that the most supersymmetric configurations are a useful, if not essential, starting point. Drawing inspiration from that body of work, the natural next step would be to consider the action on our half-BPS geometry of the symmetry generators of the $\mathcal{N}=4$ theory -- including Virasoro, R-charge and superconformal generators -- embedding the resulting solution in a more general ansatz and then using some inspired guessing to extend the solutions of that ansatz to 1/16 BPS states that are not global descendants of half-BPS ones. The states generated in this way are examples of ``multi-graviton" states and represent the fraction of the black hole ensemble accessible to classical supergravity. The formal analogy of the asymptotically AdS$_5$ microstate we have constructed with the AdS$_3$ ones gives us hope that this general program could lead to the construction of a superstratum-like family of microstates for the Gutowski-Reall black hole. It is clear, looking at the symmetries preserved by 1/16 BPS states, that the restricted truncation used in this article, \eqref{eq:truncsmall}, will not be able to accomodate states with that small amount of supersymmetry, but hopefully the more general truncation of \cite{Cvetic:2000nc} will be enough to contain at least a subset of the states. 

Leaving this program for the future, the best we can do now is to compare the half-BPS solution we have constructed with the singular geometry representing the ensemble of half-BPS states. As we have anticipated in the introduction, this is the so called superstar \cite{Myers:2001aq}, whose geometry was first found in \cite{Behrndt:1998ns,Behrndt:1998jd} (see \cite{Simon:2011zz} for a nice review of the statistical interpretation of the superstar). The 10D metric of the superstar can be written as
\begin{equation}\label{eq:superstar}
ds^2 = \sqrt{\gamma}\left[-\frac{f}{H}  \,d\tau^2 + \frac{dr^2}{f} + r^2 \,d\Omega_3^2 \right] +\sqrt{\gamma} \,d\theta^2 + \frac{H}{\sqrt{\gamma}}\cos^2\theta \, [d\phi+(H^{-1}-1) d\tau]^2+\frac{\sin^2\theta}{\sqrt{\gamma}} \,d\tilde \Omega_3^2\,,
\end{equation}
with
\begin{equation}
H=1+\frac{q}{r^2}\quad,\quad f = 1+ r^2 H\quad , \quad \gamma = 1+ \frac{q\,\sin^2\theta}{r^2}\,.
\end{equation}
It is interesting to notice that the geometry above fits in the restricted truncation \eqref{eq:truncsmall} with\footnote{As a consistency check, one can verify that the geometric data of the superstar given in \eqref{eq:superstartrun} satisfy the supersymmetry constraints \eqref{eq:susycons}.}
\begin{equation}\label{eq:superstartrun}
\lambda = 0 \quad , \quad e^{3\mu}= H \quad , \quad \Phi = H^{-1}-1 \quad , \quad \Omega_0^2 = H^{-\frac{2}{3}}\,\frac{(f - 1)\,(1-\xi^2)}{\xi^2} \quad , \quad \Omega_1^2 = H^{-\frac{2}{3}}\,f\,(1-\xi^2)\, ,
\end{equation}
\begin{equation}
	\sqrt{1 - \xi^2}  = \tanh\left(\frac{1}{\sqrt{1+q}} \log \frac{\sqrt{1+q}+\sqrt{1+q+r^2}}{r}\right) \, .
\end{equation}
In spite of this observation, we could not find a regime of parameters, $\epsilon$ and $\xi$, in which our microstate geometry reduces to the superstar. This does not contradict the expectation that the solution we have found represents one of the microstates of the superstar. It is indeed conceivable that the geometry \eqref{eq:superstar} arises from a coarse graining process over an ensemble of regular geometries, of which the one we have constructed here represents a particular one-dimensional subfamily parametrised by $\epsilon$. The fact that under this coarse graining the function $\lambda$, which is odd under $\epsilon\to -\epsilon$, averages to zero, as implied by \eqref{eq:superstartrun}, seems consistent with this interpretation.

\section*{Acknowledgments}
We would like to thank D. Turton, A. Tyukov, N. Warner and C. Wen for useful discussions. We are particularly grateful to R. Russo for many exchanges on the general problem of black hole microstates and for valuable comments on the draft. This work is supported in part by the Italian Istituto Nazionale di Fisica Nucleare and by the Research Project F.R.A. 2022 of the University of Genoa.

\appendix
\section{Equations of motion}
\label{app:eom}
The equations of motion for the truncation \eqref{eq:truncsmall} are:

\begin{subequations}\label{eq:fulleqs}

\begin{equation}\label{eq:eoms1}
\frac{(1-\xi^2)^3}{\xi^3\,\Omega_0^4 \,\Omega_1}\partial_\xi\left(\frac{\xi^3 \,\Omega_0^2 \,\Omega_1}{1-\xi^2} \,\partial_\xi \lambda\right)+2\,\frac{1-\xi^2}{\Omega_1^2}\sinh(2\lambda)(1+\Phi)^2-2\,e^{-4\mu}\sinh(2\lambda) + 8 \,e^{-\mu}\sinh\lambda=0\,,
\end{equation}

\begin{equation}\label{eq:eoms2}
\frac{(1-\xi^2)^3}{\xi^3\,\Omega_0^4 \,\Omega_1}\partial_\xi\left(\frac{\xi^3 \,\Omega_0^2 \,\Omega_1}{1-\xi^2} \,\partial_\xi \mu\right)+\frac{1}{3}\,\frac{(1-\xi^2)^3}{\Omega_0^2\,\Omega_1^2}\,e^{4\mu}(\partial_\xi\Phi)^2+\frac{4}{3}\left(e^{-4\mu}\,\sinh^2\lambda -e^{-\mu}\,\cosh\lambda+e^{2\mu}\right)=0\,,
\end{equation}

\begin{equation}\label{eq:eoms3}
\frac{\Omega_1}{\xi^3\,\Omega_0^2} \partial_\xi\left(\frac{\xi^3 \,\Omega_0^2}{\Omega_1}\,e^{4\mu}\,\partial_\xi\Phi\right)-4\,\frac{\Omega_0^2}{(1-\xi^2)^2}\,\sinh^2\lambda\,(1+\Phi)=0\,,
\end{equation}

\begin{equation}\label{eq:eoms4}
\begin{aligned}
&\frac{(1-\xi^2)^{\frac{1}{2}}}{\xi^3\,\Omega_0}\partial_\xi\left(\frac{\xi^3}{(1-\xi^2)^{\frac{1}{2}}}\,\partial_\xi\Omega_0\right)+2\frac{1-\Omega_0^2}{(1-\xi^2)^2}+\frac{1}{12}(\partial_\xi\lambda)^2+\frac{1}{2}(\partial_\xi\mu)^2\\
&+\frac{1}{3}\frac{\Omega_0^2}{(1-\xi^2)\,\Omega_1^2}\,\sinh^2\lambda\,(1+\Phi)^2+\frac{1}{12} \,\frac{1-\xi^2}{\Omega_1^2}\,e^{4\mu} (\partial_\xi\Phi)^2+\frac{1}{6}\,\frac{\Omega_0^2}{(1-\xi^2)^2}(V+12)=0\,,
\end{aligned}
\end{equation}

\begin{equation}\label{eq:eoms5}
\begin{aligned}
&\frac{1-\xi^2}{\xi^2 \,\Omega_0\,\Omega_1}\partial_\xi\left(\frac{\xi^2\,\Omega_0}{1-\xi^2}\,\partial_\xi \Omega_1\right)-\frac{(1-\xi^2)^{\frac{1}{2}}}{\xi^2\,\Omega_0^2}\partial_\xi\left(\frac{\xi^2\,\Omega_0}{(1-\xi^2)^{\frac{1}{2}}}\right)\partial_\xi\Omega_0+2\,\frac{1-\Omega_0^2}{(1-\xi^2)^2}+\frac{1}{12}(\partial_\xi\lambda)^2+\frac{1}{2}(\partial_\xi\mu)^2\\
&-\frac{5}{3}\frac{\Omega_0^2}{(1-\xi^2)\,\Omega_1^2}\,\sinh^2\lambda\,(1+\Phi)^2-\frac{5}{12} \,\frac{1-\xi^2}{\Omega_1^2}\,e^{4\mu} (\partial_\xi\Phi)^2+\frac{1}{6}\,\frac{\Omega_0^2}{(1-\xi^2)^2}(V+12)=0\,,
\end{aligned}
\end{equation}

\begin{equation}\label{eq:firstordercons}
\begin{aligned}
&\frac{(1-\xi^2)}{\xi\,\Omega_0^2} \partial_\xi\left( \frac{\xi\,\Omega_0}{1-\xi^2}\right)\partial_\xi\Omega_0+\frac{\partial_\xi(\Omega_0\,\Omega_1)}{\xi(1-\xi^2)\Omega_0\,\Omega_1}+\frac{\partial_\xi\Omega_0 \partial_\xi\Omega_1}{\Omega_0\,\Omega_1}+2\,\frac{1-\Omega_0^2}{(1-\xi^2)^2}-\frac{1}{12}(\partial_\xi\lambda)^2-\frac{1}{2}(\partial_\xi\mu)^2\\
&-\frac{1}{3}\frac{\Omega_0^2}{(1-\xi^2)\,\Omega_1^2}\,\sinh^2\lambda\,(1+\Phi)^2+\frac{1}{12} \,\frac{1-\xi^2}{\Omega_1^2}\,e^{4\mu} (\partial_\xi\Phi)^2+\frac{1}{6}\,\frac{\Omega_0^2}{(1-\xi^2)^2}(V+12)=0\,.
\end{aligned}
\end{equation}

\end{subequations}

The equations linearised around AdS$_5\times S^5$ are obtained by keeping only the terms of first order in $\lambda$, $\mu$, $\Phi$, $\omega_0\equiv \Omega_0-1$ and $\omega_1\equiv \Omega_1-1$:

\begin{subequations}\label{eq:waveops}

\begin{equation}
\nabla^2 \,\lambda\equiv \frac{(1-\xi^2)^3}{\xi^3}\partial_\xi\left(\frac{\xi^3}{1-\xi^2} \,\partial_\xi \lambda\right)+4\,(1-\xi^2)\,\lambda + 4 \,\lambda=0\,,
\end{equation}

\begin{equation}\label{eq:muhom}
\nabla^2 \,\mu\equiv \frac{(1-\xi^2)^3}{\xi^3}\partial_\xi\left(\frac{\xi^3}{1-\xi^2} \,\partial_\xi \mu\right)+4\,\mu=0\,,
\end{equation}

\begin{equation}
\nabla^2 \,\Phi \equiv \xi^{-3} \partial_\xi\left(\xi^3\,\partial_\xi\Phi\right)=0\,,
\end{equation}

\begin{equation}
\nabla^2\, \omega_0\equiv \frac{(1-\xi^2)^{\frac{1}{2}}}{\xi^3}\partial_\xi\left(\frac{\xi^3}{(1-\xi^2)^{\frac{1}{2}}}\,\partial_\xi\omega_0\right)-4\,\frac{\omega_0}{(1-\xi^2)^2}=0\,,
\end{equation}

\begin{equation}
\nabla^2\,\omega_1\equiv \frac{1-\xi^2}{\xi^2}\partial_\xi\left(\frac{\xi^2}{1-\xi^2}\,\partial_\xi \omega_1\right)=\frac{2-\xi^2}{\xi(1-\xi^2)}\partial_\xi\omega_0+4\,\frac{\omega_0}{(1-\xi^2)^2}\,.
\end{equation}

\end{subequations}

\section{Holographic computations}
\label{app:holo}
Here we provide the derivation of the holographic results presented in Section~\ref{sec:holo} for the R-charge, \eqref{eq:R-charge}, and for the expectation values of the dimension two and four CPO's,  \eqref{eq:vevo2}-\eqref{eq:vevo20}. The computations are based on the gauge-invariant formalism developed in \cite{Skenderis:2007yb}, which we summarise below to make the presentation self-consistent. 

The holographic data are encoded in the asymptotic ($\xi\to 1$) expansion of the solution, where the geometry tends to $AdS_5 \times S^5$ plus a fluctuation; we denote by $h_{MN}$, with $M=\{\mu,\alpha\}$, the components of the metric fluctuation and by $f^{(5)}$ the fluctuation of the five-form. For the expectation values of the scalar CPO's we only need the components of the metric and the 5-form along the $S^5$, which can be expanded in spherical harmonics as
\begin{equation}
	\begin{aligned}
		h_{(\alpha\beta)} &= \sum_{k,I} \phi^{(I)}_{k(t)}\, Y^I_{k(\alpha\beta)} + \phi^{(I)}_{k(v)}\,D_{(\alpha}Y^I_{k\,\beta)} + \phi^{(I)}_{(s)}\,D_{(\alpha}D_{\beta)}Y^I_k \quad,\quad h_\alpha^\alpha = \sum_{k,I} \pi^{(I)}_k \,Y^I_k\,, \\
		{f^{(5)}}_{|_{S^5}} &= \sum_{k,I} b^{(I)}_{k(s)}\,\Lambda_{k(s)}\, \star_{S^5} Y^I_k\,. 
	\end{aligned}
\end{equation}
The round parenthesis $(\alpha\beta)$ denotes the symmetric traceless part and $Y^I_k$, $Y^I_{ka}$, $Y^I_{k(ab)}$ are scalar, vector and tensor harmonics of degree $k$ with $I$ labelling the different harmonics in the same $SO(6)$ multiplet. The coefficients $\phi^{(I)}_{k(t)}$, $\phi^{(I)}_{k(v)}$, $\phi^{(I)}_{(s)}$, $\pi^{(I)}_k$, $b^{(I)}_{k(s)}$ are functions on AdS$_5$ and their asymptotics at infinity determine the VEV's we are seeking to extract. In particular they can be taken to depend on the Fefferman-Graham coordinate $z$ and the leading term for $z \to 0$ goes like $z^k$ for the coefficients of spherical harmonics of order $k$. When we will use these coefficients in the following we will always imply to have kept only these leading terms. Since we are interested in VEV's of chiral primaries up to dimension four, we can rely on the expansion of our solution \eqref{eq:asymptotic} up the second order in $(1-\xi^2)$, which is related to $z$ by \eqref{eq:zxi}. \\
The most direct way to extract $\phi^{(I)}_{(s)}$ is to use the trasversality of vector and tensor spherical harmonics
\begin{equation}
	D^\beta D^\alpha h_{(\alpha\beta)} = 4\, \sum_{k,I}\Lambda_{k(s)}\, \left(1 + \frac{\Lambda_{k(s)}}{5}\right)\,\phi^{(I)}_{(s)}\, Y^I_k\,,
\end{equation}
where $\Lambda_{k(s)} = -k(k+4)$ is the eigenvalue of the laplacian for the scalar harmonics. As explained in \cite{Skenderis:2007yb}, in order to find the supergravity fields dual to our CPO's in a coordinate independent way we have to work with the following gauge-invariant combinations
\begin{equation}
	\begin{aligned}
		&\hat{\pi}_k^{(I)} = \pi_k^{(I)} - \Lambda_{k(s)}\, \phi_{k(s)}^{(I)} \quad,\quad \hat{b}_k^{(I)} = b_{(s)k}^{(I)} - \frac{1}{2}\, \phi_{k(s)}^{(I)}\,,\\  
	\end{aligned}
\end{equation}
in term of which chiral primaries are dual to the scalar fields
\begin{equation}
	\hat{s}_k^{(I)} = \frac{1}{20\,(k+2)}[\hat{\pi}_k^{(I)} - 10\,(k+4)\,\hat{b}_k^{(I)}] \hspace{3mm} \text{with} \hspace{3mm} k \geq 2\,.
\end{equation}
The expectation values of CPO's (as well as those of their descendants with R-charge $I$) up to dimension four in terms of these fields are given by \cite{Skenderis:2007yb}
\begin{equation}\label{eq:scalarvevs}
	\begin{aligned}
		\langle \mathcal{O}_{2, I} \rangle &= \frac{8N^2}{\sqrt{3}\pi^{7/2}}\,\hat{s}^{(I)}_2 \, ,\\
		\langle \mathcal{O}_{4,I} \rangle &= \frac{N^2}{2\pi^2}\frac{4\sqrt{3}}{5}(2\,\hat{s}^{(I)}_4+\frac{2}{3}\,a_{4I,2J,2K}\,\hat{s}^{(J)}_2\,\hat{s}^{(K)}_2)\,,
	\end{aligned}
\end{equation}
where $a_{k_1I,k_2J,k_3K}$, with $I = J + K$, is the triple overlap between scalar harmonics
\begin{equation}
	a_{k_1I,k_2J,k_3K} = \int_{S^5}Y^{I}_{k_1}\, Y^{J}_{k_2}\, Y^{K}_{k_3}\,.
\end{equation}
By using the usual parametrisation for $S^5$, \eqref{eq:AdS5andS5}, the relevant scalar spherical harmonics are
\begin{equation}
	\begin{aligned}
		&Y^{0}_4 = \sqrt{\frac{3}{\pi^3}}(10 \,\cos^4\theta - 8\,\cos^2\theta + 1) \quad,\quad Y^{0}_2 = \sqrt{\frac{2}{\pi^3}}(3\,\cos^2\theta -1)\,,\\ 
		&Y^{\pm 4}_4 = \sqrt{\frac{15}{\pi^3}}\,\cos^4\theta\, e^{\pm i 4  \tilde\phi} \quad , \quad Y^{\pm 2}_2 = \sqrt{\frac{6}{\pi^3}}\,\cos^2\theta\, e^{\pm i 2  \tilde\phi}\,,
	\end{aligned} 
\end{equation}
which are unit normalized in contrast with the conventions of \cite{Skenderis:2007yb}. The triple overlaps that we need are
\begin{equation}
	a_{40,20,20} = \frac{3\sqrt{3}}{5 \sqrt{\pi^3}} \quad,\quad a_{40,22,2-2} = a_{40,2-2,22} = \frac{\sqrt{3}}{5\sqrt{\pi^3}} \quad,\quad a_{44,22,22} = a_{4-4,2-2,2-2} = 2\sqrt{\frac{3}{5\pi^3}}\,.
\end{equation}
By projecting the fluctuations determined by \eqref{eq:asymptotic} into the scalar spherical harmonic basis and constructing the gauge-invariant quantities, one finds 
\begin{equation}
	\begin{aligned}
		&\hat{s}_2^{(0)} = \frac{11}{36}\,\sqrt{\frac{\pi^3}{2}}\,(-1+\cosh\epsilon) \quad, \quad \hat{s}_2^{(\pm2)} = -\frac{11}{24}\,\sqrt{\frac{\pi^3}{6}}\,\sinh\epsilon\, , \\
		&\hat{s}_4^{(0)} = \frac{121}{17280}\,\sqrt{\frac{\pi^3}{3}}\,(-5+8\,\cosh\epsilon -3\,\cosh(2\epsilon)) \quad, \quad \hat{s}_4^{(\pm4)} = -\frac{121}{1728}\,\sqrt{\frac{\pi^3}{15}}\,\sinh^2\epsilon\, .
	\end{aligned}
\end{equation}
Using \eqref{eq:scalarvevs}, the expectation values of the neutral and maximally charged components of the dimension-two CPO are
\begin{equation}
	\langle \mathcal{O}_{2,0}\rangle = \frac{N^2}{2\pi^2} \frac{44\sqrt{2}}{9\sqrt{3}} \,\sinh^2\frac{\epsilon}{2}\, \quad , \quad \langle \mathcal{O}_2 \rangle = -\frac{N^2}{2\pi^2} \frac{11\sqrt{2}}{9} \,\sinh\epsilon\,,
\end{equation}
while those of the dimension-four CPO vanish (the other R-charge components are trivially zero, as they do not appear in the expansion of our solution)
\begin{equation}
	\langle \mathcal{O}_{4,0}\rangle = \langle \mathcal{O}_4\rangle = 0\,.
\end{equation} 
Note that this is the result of a non-trivial cancelation between the two terms in the second line of \eqref{eq:scalarvevs}.

To extract the R-charge we need the following components of the metric and five-form fluctuations:
\begin{equation}
	\begin{aligned}
		h_{\mu \alpha} &= \sum_{k,I}B_{k(v)\mu}^{(I)}\,Y^I_{k \,\alpha} + B_{k(s)\mu}^{(I)}\,D_\alpha Y^I_k\,, \\
		f_{\alpha\beta\gamma\delta\mu} &={\epsilon_{\alpha\beta\gamma\delta}}^\epsilon \sum_{k,I}D_\mu b_{k(s)}^{(I)}\, D_\epsilon Y^I_k+(\Lambda_{k(v)}-4)\,b^{(I)}_{k\mu}\,Y_{k\,
	\epsilon}^I\,,
	\end{aligned}
\end{equation}
where $\Lambda_{k(v)} = -(k^2+4k-1)$ is the eigenvalue of the laplacian for the vector harmonics. The only (unit normalized ) vector spherical harmonic relevant for our purposes is
\begin{equation}
	Y_1 = \sqrt{\frac{3}{\pi^3}}\,\cos^2\theta\, d\tilde\phi \,.
\end{equation}
From the components along this harmonic one can construct the gauge-invariant vector combination
\begin{equation}
	a_{1\tau} = B_{1(v)\tau} - 16\,b_{1\tau}\,,   
\end{equation}
and compute the R-charge using
\begin{equation}
	J = \int_{S^3}J_\tau\quad\text{with}\quad J_\tau = -\frac{N^2}{4\sqrt{3}\pi^{7/2}}\,a_{1\tau}\,.
\end{equation}
The leading terms of $B_{1(v)\tau}$, $b_{1\tau}$ go like $z^2$ for $z \to 0$ so we can use the asymptotic expansion of our solution \eqref{eq:asymptotic} at the first order in $(1-\xi^2)$. By projecting the fluctuations of our solution onto the vector harmonic $Y_1$ we find
\begin{equation}
	a_{1\tau} = -2\sqrt{3 \pi^3}\,\sinh^2\left(\frac{\epsilon}{2}\right) \quad \Rightarrow \quad J = N^2 \sinh^2\left(\frac{\epsilon}{2}\right)\,.
\end{equation}


\providecommand{\href}[2]{#2}\begingroup\raggedright\endgroup

\end{document}